\documentclass[11pt,a4]{article}
\usepackage{amsmath}
\usepackage{amscd}
\usepackage{array}
\usepackage{latexsym,amssymb}
\DeclareFontFamily{U}{rsf}{}
\DeclareFontShape{U}{rsf}{m}{n}{
  <5> <6> rsfs5 <7> <8> <9> rsfs7 <10-> rsfs10}{}
\DeclareMathAlphabet\Scr{U}{rsf}{m}{n}

\newcommand{\nn}{\nonumber}

\def\to{\rightarrow}

\newcommand{\eq}[1]{(\ref{#1})}

\newcommand{\be}{\begin{equation}}
\newcommand{\ee}{\end{equation}}
\newcommand{\bea}{\begin{eqnarray}}
\newcommand{\eea}{\end{eqnarray}}

\newcommand{\ba}{\begin{eqnarray}}
\newcommand{\ea}{\end{eqnarray}}

\def\e{\epsilon}

\def\cE{\mathcal{E}}

\def\cG{\mathcal{G}}
\def\cH{\mathcal{H}}

\def\cK{\mathcal{K}}
\def\cL{\mathcal{L}}
\def\cM{\mathcal{M}}

\def\cR{\mathcal{R}}

\def\cU{\mathcal{U}}

\def\cW{\mathcal{W}}

\def\ii{\mathrm{i}}

\def\notin{\hbox{{$\in$}\kern-.51em\hbox{/}}}

\def\inbar{\vrule height1.5ex width.4pt depth0pt}
\def\IB{\relax{\rm I\kern-.18em B}}
\def\IC{\relax\,\hbox{$\inbar\kern-.3em{\rm C}$}}
\def\ID{\relax{\rm I\kern-.18em D}}
\def\IE{\relax{\rm I\kern-.18em E}}
\def\IF{\relax{\rm I\kern-.18em F}}
\def\IG{\relax\,\hbox{$\inbar\kern-.3em{\rm G}$}}
\def\IH{\relax{\rm I\kern-.18em H}}
\def\II{\relax{\rm I\kern-.17em I}}
\def\IK{\relax{\rm I\kern-.18em K}}
\def\IL{\relax{\rm I\kern-.18em L}}
\def\IN{\relax{\rm I\kern-.18em N}}
\def\IP{\relax{\rm I\kern-.18em P}}
\def\IQ{\relax\,\hbox{$\inbar\kern-.3em{\rm Q}$}}
\def\IR{\relax{\rm I\kern-.18em R}}
\def\IU{\relax\,\hbox{$\inbar\kern-.3em{\rm U}$}}
\def\ZZ{\relax\ifmmode\mathchoice{\hbox{\cmss
Z\kern-.4em Z}}{\hbox{\cmss Z\kern-.4em Z}}{\lower.9pt\hbox{\cmsss Z\kern-.4em Z}} {\lower1.2pt\hbox{\cmsss
Z\kern-.4em Z}}\else{\cmss Z\kern-.4em Z}\fi}
\def\IGam{\relax{{\rm I}\kern-.18em \Gamma}}

\def\bfnull{\relax{\rm O \kern-.635em 0}}

\def\square{{\,\lower0.9pt\vbox{\hrule
\hbox{\vrule height 0.2 cm \hskip 0.2 cm \vrule height 0.2 cm}\hrule}\,}}

\def\twomat#1#2#3#4{\left(\begin{array}{cc} \end{array} \right)}

\begin{document}

\numberwithin{equation}{section}

\begin{center}
{\bf\LARGE Rotating black holes, global symmetry and \\first order formalism} \\
\vskip 2 cm
{\bf \large Laura Andrianopoli, Riccardo D'Auria, \\
 Paolo Giaccone and Mario Trigiante}
\vskip 8mm
 \end{center}
\noindent {\small{\it DISAT, Politecnico di Torino, Corso Duca
    degli Abruzzi 24, I-10129 Turin, Italy and Istituto Nazionale di
    Fisica Nucleare (INFN) Sezione di Torino, Italy}

\vskip 2 cm
\begin{center}
{\small {\bf Abstract}}
\end{center}
In this paper we consider axisymmetric black holes in supergravity and  address the general issue
of defining a first order description for them. The natural setting where to formulate  the problem is the
De Donder--Weyl--Hamilton--Jacobi theory associated with the effective two-dimensional sigma-model action
describing the axisymmetric solutions. We write the general form of the two functions $S_m$ defining the first-order equations for the fields.
 It is invariant under the global symmetry group $G_{(3)}$ of the sigma-model.
 We also discuss the general properties of the solutions with respect to these global symmetries, showing that they can be encoded in two constant matrices belonging to the Lie algebra of $G_{(3)}$, one being the N\"other matrix of the sigma model, while the other is non-zero only for rotating solutions.
These two matrices allow a $G_{(3)}$-invariant characterization of the rotational properties of the solution and of the extremality condition. We also comment on extremal, under-rotating solutions from this point of view.
\vskip 1 cm
\vfill
\noindent {\small{\it
    E-mail:  \\
    {\tt
      laura.andrianopoli@polito.it}; \\
      {\tt riccardo.dauria@polito.it}; \\
      {\tt p.giaccone@polito.it}; \\
      {\tt mario.trigiante@polito.it}}}
   \eject

%%%%%%%%%%%%%%%%%%%%%%%%%%%%%%%%%%%%%%%%%%%%%%%%%%%%%
%%%%%%%%%%%%%%%%%%%%%%%%%%%%%%%%%%%%%%%%%%%%%%%%%%%%%
%%%%%%%%%%%%%%%%%%%%%%%%%%%%%%%%%%%%%%%%%%%%%%%%%%%%%

%\begin{center}
%\today
%\end{center}

%%%%%%%%%%%%%%%%%%%%%%%%%%%%%%%%%%%%%%%%%%%%%%%%%%%%

\section{Introduction}
\label{intro} There has been a considerable progress in the knowledge of static black holes in supergravity,
both from the point of view of finding solutions  and of their classification \cite{reviews,reviews2}, in four and higher
dimensions.

A relevant role in these developments was played by the use of a first order formalism,
corresponding to the introduction  of a fake-superpotential \cite{Ceresole:2007wx,Lopes Cardoso:2007ky,Andrianopoli:2007gt,Bellucci:2008sv,Andrianopoli:2009je,Ceresole:2009iy,Bossard:2009we,Ceresole:2009vp,Andrianopoli:2010bj,Andrianopoli:2011zz} that was recognized to be
strictly related to the Hamilton characteristic function in a mechanical problem where the evolution is in the
radial variable $\tau$ \cite{Andrianopoli:2007gt,Andrianopoli:2009je,Andrianopoli:2010bj}. The latter approach naturally applies to both extremal and  non-extremal static, single center black holes.

As far as more general solutions, such as  stationary and/or multicenter black holes \cite{Behrndt:1997ny,Denef:2000nb,Gauntlett:2002nw,LopesCardoso:2000qm,Bena:2009ev,Dall'Agata:2010dy}, are concerned, a similar
comprehensive study is still missing. In particular, the use of a first order formalism has not been much
exploited except in very particular cases \cite{Galli:2010mg,Yeranyan:2012au,Bossard:2012xs}.

A peculiarity of static, spherically symmetric solutions is that one can exploit the symmetries to reduce the
Lagrangian to a one-dimensional effective one, where the evolution variable is the radial one
\cite{Gibbons:1996af,Ferrara:1997tw}. However, when considering four dimensional solutions with less symmetries, in particular
stationary solutions where only the time-like Killing vector $\partial_t$ is present, an effective
three-dimensional Lagrangian can be obtained upon compactification along the time coordinate
\cite{Breitenlohner:1987dg,Gunaydin:2007bg,Gaiotto:2007ag,Bergshoeff:2008be,Bossard:2009at,Chemissany:2010zp,Fre:2011uy,Bossard:2012ge,Chemissany:2012nb,Fre:2012im}. The fields in the effective Lagrangian now depend on the three space variables
$x^i$, ($i=1,2,3$). In particular, for stationary axisymmetric solutions, the presence of an azimuthal angular
Killing vector $\partial_\varphi$ allows a further dimensional reduction to two dimensions.

 The problem of extending the Hamilton--Jacobi (in the following, HJ) formalism from mechanical models, whose degrees of freedom depend on just one
 variable, to field theories where the degrees of freedom depend on two or more variables, was addressed and
 developed in generality from several points of view (a useful review is given by \cite{Kastrup:1982qq}).

Our main aim in the present paper is to apply such extended  formalism in the study of black holes. We
 will adhere to the so-called De Donder--Weyl--Hamilton--Jacobi theory, hereafter referred to as DWHJ, which is the simplest extension of the
classical HJ approach in mechanics. One important difference with respect to the case of classical mechanics
consists in the replacement of the Hamilton principal function $S$ (directly related to the fake-superpotential
of static black holes) with a \emph{Hamilton principal $1$-form}, that is with a covariant vector $S_i$.

As it  is usual in the three dimensional approach, by using Hodge-duality in three dimensions all the fields of
the parent four dimensional theory are described by three dimensional scalars \cite{Breitenlohner:1987dg}
and their interaction is given by gravity coupled to a $\sigma$-model. Correspondingly, the equations of motion
give a set of conserved currents. A particularly interesting case is when the $\sigma$-model is a
symmetric space $G_{(3)}/H^*$ (where $H^*$ denotes a suitable non-compact maximal subgroup of $G_{(3)}$ \cite{Breitenlohner:1987dg}). Note that the effective geodesic Lagrangian is invariant under the
three-dimensional isometry group $G_{(3)}$ (we will also refer to it as the three-dimensional \emph{duality
group}). One of the main results of our paper is  to give a manifestly duality invariant expression for the
Hamilton principal vector $S_i$, thus extending the results obtained for the Hamilton characteristic function
$\cW$ in the static case \cite{Andrianopoli:2009je}.

For pure Einstein--Maxwell stationary configurations, the three-dimensio\-nal $\sigma$-model turns out to be
$\mathrm{SU(1,2)/U(1,1)}$. As is well known in General Relativity, in the presence of a time-like Killing vector
Einstein--Maxwell theory is very efficiently described in terms of the so-called Ernst potentials  $\cE$, $\Psi$
(see for example \cite{Stephani:2003tm,Chandrasekhar:1985kt}), which are complex functions of the
$\mathrm{SU(1,2)}$ complex triplet of fields $\mathbb{U}=(W,V,U)$. We found particularly useful, outside the
ergosphere, to parametrize the coset $\mathrm{SU(1,2)/U(1,1)}$ with the homogeneous fields $U,V,W$, or more
precisely with their inhomogeneous counterpart $(u=U/W,v=V/W)$, corresponding to four real scalar degrees of
freedom.

In the present paper we will give general results on stationary axisymmetric solutions of four dimensional
supergravity and then
 focus on the first-order formulation of the Kerr-Newman solution and its extension in the presence of a
NUT charge. Besides finding a duality invariant $S_i$, we will also express the conserved charges of the
black hole \cite{Wald} in terms of the conserved charges of the $\sigma$-model $G_{(3)}/H^*$. Actually, the
N\"other charges associated with $G_{(3)}$ global symmetry  do not include the angular momentum $M_\varphi$. The
latter can nevertheless be expressed in terms of quantities which are intrinsic to the $\sigma$-model. This is
achieved by introducing a new
$G_{(3)}$-covariant constant matrix, besides the N\"other charge one $Q$, defined as follows:
\begin{equation}
Q_\psi =-\frac{3}{8\pi}\,\int_{S_2^\infty}\psi_{[i}\,J_{
j]}\,dx^i\wedge dx^j\,,
\end{equation}
$J_i$ being the N\"other current with value in the algebra of $G_{(3)}$ and $\psi=\partial_\varphi$ the
azimuthal angle Killing vector.
From straightforward application of the general four-dimensional expression for
the angular momentum one finds that its squared value, for the Kerr-Newmann solution, can be written as the ratio of two $G_{(3)}$ invariants
$\mathrm{Tr}(Q^2_\psi)$ and $\mathrm{Tr}(Q^2)$, and thus can be given a description which is invariant with
respect to the global symmetry of the $\sigma$-model and is straightforwardly generalizable to more general
 models with $D=4$ scalar fields. This analysis  also provides a $G_{(3)}$-invariant characterization of the extremality parameter (and thus of the extremality condition), see eq.s (\ref{eqq3'}), (\ref{eqq3}), so that the cosmic-censor condition for Kerr black holes, $M_{ADM}^4\geq M_\varphi^2$, can be recast for the generic regular axisymmetric solution, in  a $G_{(3)}$-invariant way as
 $$[\mathrm{Tr} (Q^2)]^2\geq \frac 2k\,\mathrm{Tr} (Q_\psi^2)\,,$$
  $k$ being a representation-dependent constant. In particular we show that in the  extremal ``ergo-free'' solutions \cite{Horne:1992zy,Rasheed:1995zv,Larsen:1999pp,Astefanesei:2006dd,Bena:2009ev}, both matrices $Q,\,Q_\psi$ are \emph{nilpotent}, the former having a larger degree of nilpotency of the latter. The first-order formalism and the functions $S_m$ for for under-rotating solutions were derived in \cite{Yeranyan:2012au}.

 A description of the global symmetry properties of  axisymmetric solutions should then
 include at least the \emph{two independent, mutually orthogonal matrices $Q,\,Q_\psi$} belonging to the Lie
 algebra of the global symmetry group.

The paper is organized as follows: \\In Section \ref{hj} we present the extension of the HJ theory to field
theory, following the DWHJ approach, and give a general formula to find the Hamilton principal 1-form. \\In section
\ref{axi} we focus on stationary axisymmetric black holes, whose description, following
\cite{Breitenlohner:1987dg}, is two dimensional. We review the construction of the two-dimensional effective Lagrangian and the expression of the characteristic physical quantities associated with the four-dimensional solution in terms of N\"other currents of the sigma-model.
We also write the angular momentum in terms of the sigma-model N\"other currents and introduce, besides $Q$, the matrix $Q_\psi$, which allows to describe in a $G_{(3)}$-invariant fashion the rotational properties of the solution. We also discuss the \emph{under-rotating} extremal limit of a non-extremal solution in the $G_{(3)}$-orbit of the Kerr-black hole.
Then we find a
manifestly (three-dimensional) duality invariant expression for the principal functions $S_m$ ($m=1,2$). \\In
Section \ref{universal} we restrict our attention to the KN-Taub-NUT solution, making use of the Ernst
potentials written in terms of the inhomogeneous fields $(u,v)$ to parametrize the $\mathrm{SU(1,2)/U(1,1)}$
coset and give the explicit form of the principal functions $S_m$ in terms of fields and two-dimensional coordinates.\\
We end in Section \ref{concl} with some concluding remarks. \\ Appendix \ref{appe}
contains the explicit form of the algebra $\mathrm{SU(1,2)}$, while Appendix \ref{symcle}, extending the procedure of \cite{Clement:1997tx} to the case where a NUT charge is present, shows how one can
retrieve the KN-Taub-NUT solution from Schwarzschild by use of duality and general coordinate
transformations. In particular,
Appendix \ref{symcle1} contains a manifestly $H^*$-invariant expression for the $\mathcal{W}$-function
describing the RN-Taub-NUT solution in the universal model, which, to our knowledge, was not known so far, and then applies a known procedure \cite{Clement:1997tx} to generate from it the KN-Taub-NUT solution by a set of duality and general coordinate transformations.

\section{Hamilton--Jacobi formalism for field theory}\label{hj}
In a previous work a formalism was developed to interpret the first-order description of static black holes in
terms of Hamilton--Jacobi theory. In particular, the Hamilton characteristic function $\cW$ was shown to be related, for
extremal solutions, to the ``fake'' superpotential: $\cW= 2\,e^{2U} W$
\cite{Andrianopoli:2007gt,Andrianopoli:2009je}.  The above
construction works well in the static, spherically symmetric case where the metric only depends in a non-trivial
way on the evolution radial variable $\tau$ so that the Einstein Lagrangian can be reduced to an effective
one-dimensional Lagrangian. For more general black holes, with a lower number of isometries, we have to extend
the Hamilton--Jacobi formalism to a more general setting. In particular, for stationary
  black holes corresponding to the existence of a
Killing vector  associated  with time translations
$\frac{\partial}{\partial t}$, the metric can be reduced to
the following  general form
\begin{equation}
ds^2= e^{2\cU}(dt + \omega)^2 - e^{-2\cU}g_{ij}dx^i dx^j
\end{equation}
where  the fields $\cU,\omega=\omega_i dx^i $ and the 3D metric  tensor $g_{ij}$ depend on the space coordinates
$x^i$, $i=1,2,3$.

In the static, spherically symmetric case, the HJ equations arise in a classical mechanical effective model where the evolution
variable $\tau$ plays the role of time.
 A first-order formulation for a more general black-hole solution requires the extension of the Hamilton--Jacobi description from
classical mechanics  to a field theory depending on two or more variables (see, for example, \cite{Kastrup:1982qq} and
references therein). In this setting the Hamilton--Jacobi description has to be generalized to the so-called De
De Donder--Weyl--Hamilton--Jacobi theory, hereafter referred to as DWHJ, which amounts to the following. Let
$\cL(z^a, v^a_i,x^i)$ be the Lagrangian density of the system, where $z^a$ ($a=1,\cdots ,n$) are the field
variables which become functions of the $x^i$, $z^a=\xi^a(x)$, on the extremals, while $v^a_i =
\partial_i \xi^a$ on the extremals.\footnote{With an abuse of notation, we will often use $\partial_i z^a$ to denote the $v^a_i$.} The canonical momenta are defined by $\pi^i_a = \frac{\partial \cL}{\partial
v^a_i}$, and the invariant Hamilton density function is
\begin{equation}
\cH= \pi^i_a v^a_i -\cL\,.\label{hl}
\end{equation}
 The DWHJ
equation is a first-order partial differential equation for the functions $S^i(z,x)$:
\begin{equation}
\partial_i S^i (z,x) + \cH (z,x,\pi)=0\,,\label{hj0}
\end{equation}
where
\begin{equation}
\pi^i_a = \partial_a S^i(z,x)\,.
\end{equation}
The functions $S_i =\frac1{\sqrt{g}} g_{ij} S^j$   may be thought of as the components of a one-form
$S^{(1)}\equiv S_i dx^i$.\footnote{We observe that, in the presence of a gravitational field, which is the case
we will deal with, \eq{hj0} should be modified to contain the covariant divergence $\nabla_i S^i$. However,
defining the contravariant vector density  $S^i \equiv \sqrt{g} g^{ij} S_j$, $S_j$ being a true covariant vector, makes
 it possible to trade the covariant derivatives for ordinary ones,
 so that the equations are formally the same as in flat space. In
 this case, however, by $\cal H$ we mean the hamiltonian density including the factor $\sqrt{|g|}$.
 }

 In the field-theory case the issue of integrability is more involved than in mechanics since, even if a
complete integral $S^i$ can be found, solutions to the Euler--Lagrange equations can be
 constructed if the integrability conditions (which are trivial in mechanics):
\begin{equation}
\partial_{[i} v^a_{j]}=0 \label{integ}
\end{equation}
are satisfied. Taking into account that $v^a_i(\pi,z,x)= v^a_i\left(\frac{\partial S}{\partial z},z,x\right)$,
this imposes severe constraints on the solutions $S^i(z,x)$.  From now on we will mainly focus on the two
dimensional case, which is  relevant when discussing axisymmetric black holes for which two Killing vectors
exist, associated with time translations $\frac{\partial}{\partial t}$ and rotations about an axis
$\psi=\frac{\partial}{\partial \varphi}$. Note however that the extension of the formalism from systems depending on
two independent variables to systems with three or more independent variables is straightforward and does not
bring anything conceptually new \cite{Kastrup:1982qq}. We will denote the independent variables for the
two-dimensional case by $x^m$, $m=1,2$. The 3D metric in this case takes the form: $g_{ij} dx^i dx^j =
\gamma_{mn}dx^m dx^n + \hat\rho^2 d \varphi^2$, where $\varphi$ denotes the azimuthal angle about the rotation
axis, and  the fields $\gamma_{mn},\hat\rho$ depend on $x^m$.

If one introduces the two-form Lagrangian
\begin{eqnarray}
\Omega_0 &=& -\cH dx^m\wedge dx^n +\pi^m_a d\xi^a \wedge
dx^n\epsilon_{mn}
\end{eqnarray}
then the Hamilton--Jacobi equations are given by the condition
\begin{equation}
d \Omega_0 =0
\end{equation}
which implies that, locally, there exist two functions $S^m$ in
terms of which $\Omega_0$ can be written in the following form:
\begin{eqnarray}
\Omega_0&=& dS^m \wedge dx^n\epsilon_{mn}\,,
\end{eqnarray}
so that \footnote{We denote with $\partial_m$ the derivative with respect to explicit $x^m$ dependence, while total derivative with respect
to $x^m$ is denoted by $\frac{d}{dx^m}$:
\begin{equation}
\frac{d}{dx^m}f(\xi,x)\equiv \partial_m \xi^a \frac{\partial f}{\partial \xi^a}+\partial_m f
\end{equation}
}
\begin{eqnarray}
\partial_m S^m&=&-\cH\,,
\label{hj1}\\
\frac{\partial S^m}{\partial z^a}&=& \pi^m_a\,\label{hj2}.
\end{eqnarray}

\subsection{Solving DWHJ equations} In the present section we discuss in a general setting a possible way to solve
the DWHJ equations. Then, in the next sections we will apply this procedure to the study of axisymmetric black
holes and their Taub-NUT extensions.
We will give here a constructive recipe to find solutions to the field
equations by solving the DWHJ equations, following a general procedure   given in the literature (see for
example \cite{Kastrup:1982qq} and references therein).

As already anticipated, in field theory the expression for $S^m$ is strongly restricted by the integrability
constraints (\ref{integ}). In particular, as opposed to the one-dimensional classical-mechanics case, it is not
always possible to find an expression for $S^m$ valid in an open neighborhood of the extremals $z^a = \xi^a(x)$
in the space of fields \emph{and} coordinates. When this is possible, one says that the extremals $z^a =
\xi^a(x)$ are \emph{strongly} embedded in the wave fronts $S^m(z,x)$. In many cases, however, the solution $S^m$
satisfies eqs. (\ref{hj1}) and (\ref{hj2}) only on the extremals $z^a = \xi^a(x)$. One then says that the
extremals are \emph{weakly} embedded in $S^m(z,x)$.

A possible solution which is weakly embedded in $S^m$ is found by choosing one of the $x^m$, say $x^1$, as the
evolution variable:
\begin{equation}
S^m= \left(z^a -\xi^a(x)\right)\pi^m_a(\xi,x) +
\delta^m_1 \int^{x^1} {dx^1}^\prime \cL(\xi(x'),\partial_m\xi,x')
+ \mathcal{O}[(z^a -\xi^a(x))^2]
\label{kda}
\end{equation}
Indeed, from \eq{kda} we find, using \eq{hl}
\begin{eqnarray}
\partial_a S^m |_{z=\xi}&=& \pi^m_a\\
\partial_m S^m|_{z=\xi}&=& -\partial_m\xi^a \pi^m_a + \cL(\xi(x'),\partial_m\xi,x')= -\cH(\xi(x'),\partial_m\xi,x')\,.
\end{eqnarray}
Eq. \eq{kda} can be understood as a linear approximation of the Taylor expansion of $S^m$ in the neighborhood of the
extremal.
\section{The 2D Effective Lagrangian and its  Field-Theoretical DWHJ description}\label{axi}
In the presence of a time-like Killing vector $\partial_t$, the vielbein $V^a$ ($a=0,1,2,3$) of space-time can
be put in the form
\begin{equation}
V^0= e^\cU(dt+\omega)= e^{\cU} D^0\,;\quad V^i = e^{-\cU} D^i
\end{equation}
where $D^i$ ($i=1,2,3$) are 3D vielbein. The time-reduced 3-dimensional Lagrangian describing a stationary 4D
black hole in the presence of a given number of scalars $\phi^r$ and gauge fields $A^\Lambda$ has the following
form\footnote{For the $D=4$ supergravity theory we use the units $\hbar=c=8\pi G=1$ and the normalization of the vector fields as in \cite{reviews2}.}
\begin{align} \label{geodaction}
\frac{1}{\sqrt{g_{(3)}}}\,\mathcal{L}_{(3)} &= \frac{1}{2}\,\cR - \tfrac{1}{2}
G_{ab}(z)\partial_i{z}^a\partial^i{z}^b=\nn\\
&=\frac{1}{2}\, \cR -[ \partial_i \cU \partial^i
\cU+\tfrac{1}{2}\,G_{rs}\,\partial_i{\phi}^r\,\partial^i{\phi}^s +
\tfrac{1}{2}\e^{-2\,\cU}\,\partial_i{{\bf
Z}}^T\,\mathcal{M}_{(4)}\,\partial^i{{\bf Z}} +\nn\\ &+
\tfrac{1}{4}\e^{-4\,\cU}\,(\partial_i{a}+{\bf
Z}^T\mathbb{C}\partial_i{{\bf Z}})(\partial^i{a}+{\bf
Z}^T\mathbb{C}\partial^i{{\bf Z}})]\,,
\end{align}
where $g_{(3)}\equiv {\rm det}(g_{(3)})$.
Here, all the propagating degrees of freedom have been reduced to scalars by 3D Hodge-dualization
\cite{Breitenlohner:1987dg}. In particular, the scalars ${\bf{Z}} =
(\mathcal{Z}^\Lambda,\mathcal{Z}_\Lambda)= \{\mathcal{Z}^M\}$ include the electric components $A^\Lambda_0$ of
the 4D vector fields together with the Hodge dual of their magnetic components $A^\Lambda_i$ ($i=1,2,3$) and $a$
is related to the Hodge-dual of the 3D graviphoton $\omega_i$. More precisely,
\begin{eqnarray}
A^\Lambda_{(4)}&=& A^\Lambda_0 D^0+ A^\Lambda_{(3)} \,,\quad A^\Lambda_{(3)}\equiv A^\Lambda_i D^i\,,\\
\mathbf{F}^M_{(4)}&=&\left(
                       \begin{array}{c}
                         F^\Lambda_{(4)} \\
                         \cG_{\Lambda (4)} \\
                       \end{array}
                     \right)
                     =d \mathcal{Z}^M \wedge D^0 + e^{-2 \cU}\mathbb{C}^{MN}\mathcal{M}_{(4) NP} {}^* d\mathcal{Z}^P\,,\\
da&=& - e^{4 \cU} {}^* d\omega - \mathbf{Z}^T \mathbb{C} d\mathbf Z\,,
\end{eqnarray}
where $F^\Lambda_{(4)}= dA^\Lambda_{(4)}$, $\cG_{\Lambda (4)} =-\frac 12 {}^*\left(\frac{\partial \cL}{\partial
F^\Lambda_{(4)}}\right)$, and
 $\cM_{(4)}(\phi)$ is the negative-definite symmetric, symplectic matrix depending on 4D scalar fields
introduced in \cite{Ceresole:1995ca,Andrianopoli:1996ve}.

The isometry group  $G_{(3)}$ of the $\sigma$-model metric $G_{ab}(z)$ contains as non trivial subgroups the
4-dimensional U-duality group $G_{(4)}$ times the  group $SL(2,\mathbb{R})$ (the Ehlers group) under which the
degrees of freedom of the 4d metric transform. The simplest 3D model is the one originating form a pure 4D
Einstein--Maxwell gravitational theory with a single time-like Killing vector. In this case $G_{(4)}={\rm U}(1)$
and the 3D $\sigma$-model has the homogeneous-symmetric target space $\frac{{\rm SU}(1,2)}{{\rm U}(1)\times {\rm
SU}(1,1)}$. Its field content consists of four scalars belonging to a pseudo-Riemannian version of the universal
hypermultiplet, dubbed the universal \emph{pseudo-hypermultiplet}. We will discuss in more detail the properties
of this theory in the following subsection \ref{universal}.

We will mainly focus our attention on stationary axisymmetric solutions admitting the two Killing vectors
$\partial_t$ and $\partial_\varphi$. In this case one may further reduce the 3D Lagrangian to two dimensions by
compactification along $\varphi$. The fields now depend on the space coordinates   $x^m$, $m=1,2$, and we assume that the
three-dimensional space metric can be expressed in block-diagonal form as:
\begin{equation}
g_{(3)}= \left(
           \begin{array}{cc}
             \lambda^2 h_{mn} & 0 \\
             0 & \hat\rho^2 \\
           \end{array}
         \right)\,.
\end{equation}
 The resulting 2D Lagrangian
takes the form \cite{Breitenlohner:1987dg}
\begin{eqnarray} \label{geodaction2d}
\mathcal{L}_{(2)} &=&\sqrt{h}\,
\hat \rho\left(\frac{\cR_{(2)}}2- \tfrac{1}{2}
G_{ab}(z)\partial_m{z}^a\partial^m{z}^b + \frac{\partial_m\hat\rho\,\partial^m
\lambda}{\lambda\hat\rho}\right)
 \,,
\end{eqnarray}
with $h\equiv {\rm det}(h_{mn})$.
As shown in \cite{Breitenlohner:1987dg}, the dynamics of the fields $z^a$ is totally captured by the
$\sigma$-model effective action:
\begin{equation}
S_{eff}=\int d^2 x \,\sqrt{h}\,\tfrac{\hat\rho}{2}
G_{ab}(z)\partial_m{z}^a\partial^m{z}^b\,,
\end{equation}
where $\hat\rho(x^m)$ is a harmonic function in the subspace spanned by $x^m$.\footnote{According to a general
procedure in General Relativity one can perform a coordinate transformation such that the field $\hat\rho$ is
chosen as one of the new harmonic coordinates, the second coordinate $z$ being defined by $dz = -{}^\star
d\hat\rho$. Here ${}^\star$ denotes Hodge-dualization in two dimensions. In these new variables
$x^m=(\hat\rho,z)$, named Weyl-coordinates, the 2D metric is conformally flat $\gamma_{mn}= \lambda^2
\delta_{mn}$ \cite{Chandrasekhar:1985kt,Breitenlohner:1987dg}.} The metric on this space can be made conformally
flat by a suitable choice of the $x^m$ and the conformal factor absorbed in the definition of $\lambda$, so that
the equations for $z^a$ and $\hat\rho$ can be written in a flat 2D space (with ${\cR_{(2)}}=0$) spanned by
$x^m$, with metric $h_{mn}$. As we shall show in Sect. \ref{chan}, in suitable coordinates, $\sqrt{h}\,\hat{\rho}=\sin\theta$.

The equation for $\lambda$ can then be solved once the solutions to the
$\sigma$-model are known \cite{Breitenlohner:1987dg}.

We shall restrict our analysis to symmetric supergravities in which the scalar manifold $\mathcal{M}_{scal}$ of
the $D=3$ theory, spanned by the $z^a$, is homogeneous symmetric, i.e. of the form
\begin{equation}
\mathcal{M}_{scal}=\frac{G_{(3)}}{H^*}\,.
\end{equation}
We shall use for this manifold the solvable Lie algebra parametrization by identifying the scalar
fields $z^a$ with parameters of a suitable solvable Lie algebra. Let us recall the main points
\cite{Chemissany:2010zp}.
 The isometry group $G_{(3)}$ of the target space is the global symmetry group of the $S_{eff}$
and $H^*$ is a suitable non-compact semisimple maximal subgroup of it. The scalars
$z^a=\{\cU,\,a,\,\phi^r,\,\mathbf{Z}\}$ correspond to a \emph{local} solvable parametrization, i.e. the
corresponding patch, to be dubbed \emph{physical patch} ${\Scr U}$, is isometric to a solvable Lie group
generated by a solvable Lie algebra $Solv$:
\begin{equation}
\mathcal{M}_{scal}\supset {\Scr U}\equiv e^{Solv}\,,
\end{equation}
$Solv$ is defined by the Iwasawa decomposition of the Lie algebra
$\mathfrak{g}$ of $G_{(3)}$ with respect to its maximal compact
subalgebra $\mathfrak{H}$. The solvable parametrization $z^a$
can be defined by the following exponential map:
\begin{equation}
\mathbb{L}(z^a)=\exp(-a T_\bullet)\,\exp(\sqrt{2}
\mathcal{Z}^M\,T_M)\,\exp(\phi^r\,T_r)\,\exp(2\cU
T_0)\,,\label{cosetr3}
\end{equation}
where the  generators $T_0,\,T_\bullet,\,T_r,\,T_M$ satisfy the
following commutation relations:
\begin{align}
[T_0,\,T_M]&=\frac{1}{2}\,T_M\,\,;\,\,\,[T_0,\,T_\bullet]=T_\bullet\,\,;\,\,\,[T_M\,T_N]=\mathbb{C}_{MN}\,T_\bullet\,,\nonumber\\
[T_0,T_r]&=[T_\bullet,T_r]=0\,\,;\,\,\,[T_r,T_M]=T_r{}^N{}_M\,T_N\,\,;\,\,\,[T_r,T_s]=-
T_{rs}{}^{s'} T_{s'}\,,\label{relc1}
\end{align}
$T_r{}^N{}_M$ representing the symplectic representation of $T_r$ on
contravariant symplectic vectors $d\mathcal{Z}^M$. We can use for
the generators of $\mathfrak{g}$ a representation in which the
generators of $\mathfrak{H}^*$, the Lie algebra of $H^*$, are invariant
under the involution $\sigma: M\rightarrow -\eta M^\dagger \eta $,
where $\eta\equiv (-1)^{2\,T_0}$. The vielbein $P$ and connection
${\Scr W}$ 1-forms on the manifold are computed as the odd and even
components, respectively, of  the left-invariant one-form with
respect to $\sigma$:
\begin{equation}
\mathbb{L}^{-1}d \mathbb{L}=P+{\Scr W}\,,
\end{equation}
$P=\eta P^\dagger \eta=-\sigma(P)$, ${\Scr W}=-\eta {\Scr W}^\dagger
\eta=\sigma({\Scr W})$. In terms of $P$ the metric on the manifold reads:
\begin{equation}
dS^2_{(3)}=G_{ab}(z)dz^a\,dz^b=k\,{\rm Tr}(P^2)\,,\label{geo3}
\end{equation}
where $k=1/(2 {\rm Tr}(T_0^2))$ is a representation-dependent constant. It is also useful to introduce the
hermitian, $H^*$-invariant matrix $\mathcal{M}$:
\begin{equation}
\mathcal{M}(z)\equiv \mathbb{L}\eta
\mathbb{L}^\dagger=\mathcal{M}^\dagger\,,\label{cm}
\end{equation}
in terms of which we can write the geodesic Lagrangian as:
\begin{equation}
\cL_{(2)eff}= \frac 12\hat\rho \,\sqrt{h}\,G_{ab}(z)\partial_m z^a\,\partial^m z^b=
\frac k{8 }\hat\rho \,\sqrt{h}\,\mathrm{Tr}\left[\mathcal{M}^{-1}\partial_m \cM \mathcal{M}^{-1}\partial^m \cM\right]\,,\label{geolag}
\end{equation}
with a canonically conjugate momentum
\begin{equation}
\pi^m_a= \frac{\partial \cL}{\partial \partial_m z^a}=\frac k{4 }\hat\rho \,\sqrt{h}\,\mathrm{Tr}\left[\mathcal{M}^{-1}(z)\partial_a \cM (z)
\mathcal{M}^{-1}(z)\partial_b \cM (z)\right]
 \,\partial^m z^b\,. \label{pigeo}
\end{equation}
 The corresponding  equations of motion are:
\begin{equation}
\partial_m\left(\sqrt{h}\,\hat\rho h^{mn}J_n\right)=0\,,
\end{equation}
where
\begin{equation}
J_m\equiv \frac{1}{2}\partial_m\xi^a\,\mathcal{M}^{-1}\partial_a\label{curr}
\mathcal{M}\,.
\end{equation}
\subsection{Conserved quantities}
Note that the quantity $\hat \rho J=\hat\rho J_m\,dx^m$ is a 1-form N\"other current of the two-dimensional effective theory with value in $\mathfrak{g}$
implying that the integral:
\begin{equation}
Q=\frac{1}{4\pi}\int_{S_2} {}^{*_{3}}J=\frac{1}{2}\,\int \sqrt{h}\,h^{rr}\hat\rho
J_r d\theta \,,
\end{equation}
on a radius $r$ sphere $S_2$ is an $r$-independent matrix in $\mathfrak{g}$.

From it we may derive the set of N\"other currents $J_{A\,m}$ and the corresponding  \emph{constants of motion} $Q_A$ characterizing
the solution at radial infinity:
\begin{equation}
J_{A\,m}\equiv k \,{\rm Tr}\left(T_A^\dagger \,J_m\right)\,\,,\,\,\,Q_A=k\,{\rm Tr}\left(T_A^\dagger \,
Q\right)=\frac{1}{4\pi}\,\int_{S_2} {}^{*_3} J_A=\frac{1}{2}\,\int
\sqrt{h}\,\rho h^{rr}\,J_{A\,r}
d\theta\,,\label{QA}
\end{equation}
which consist in the ADM mass $m$ ($T_A=T_0$), the NUT charge $\ell$ ($T_A=T_\bullet$), the $D=4$ scalar charges
$\Sigma_r$ ($T_A=T_r$) and the electric-magnetic charges $\Gamma^M$  ($T_A=T_M$). The currents $J_{A\,m}$  read:
\begin{align}
J_{\bullet m}&=\frac{k}{2}\,{\rm Tr}(T_\bullet^\dagger
\mathcal{M}^{-1}\partial_m
\mathcal{M})=-\frac{1}{2}\,e^{-4\mathcal{U}}\,(\partial_m a  +{\bf
Z}^T\mathbb{C}\partial_m {\bf Z})\,,\nonumber\\
J_{0\,m}&=\frac{k}{2}\,{\rm Tr}(T_0^\dagger \mathcal{M}^{-1}\partial_m
\mathcal{M})=\partial_m \mathcal{U}+\frac{1}{2}\,e^{-2\mathcal{U}}\,{\bf
Z}^T\mathcal{M}\partial_m {\bf Z}-a \,J_{\bullet m}\,,\nonumber\\
J_{M\,m}&=\frac{k}{2}\,{\rm Tr}(T_M^\dagger
\mathcal{M}^{-1}\partial_m
\mathcal{M})=\frac{1}{\sqrt{2}}\,e^{-2\mathcal{U}}\,\mathcal{M}_{(4)\,MN}\,\partial_m
{\cal Z}^N+\sqrt{2}\,\mathbb{C}_{MN}\,{\cal
Z}^N\,J_{\bullet m}\,,\nonumber\\
J_{s\,m}&=\frac{k}{2}\,{\rm Tr}(T_s^\dagger \mathcal{M}^{-1}\partial_m
\mathcal{M})=\frac{1}{\sqrt{2}}\,\mathbb{L}_{4\,s}{}^{\hat{s}'}\,V_{4\,s''}{}^{\hat{s}'}\partial_m
\phi^{s''}+ e^{-2\mathcal{\mathcal{U}}}\,{\bf Z}^T T_s
\mathcal{M}_{(4)}\,\partial_m {\bf Z}-\nonumber\\&-T_{s MN}{\cal Z}^M
{\cal Z}^N\,J_{\bullet m}\,,
\end{align}
where $\mathbb{L}_{4\,s}{}^{\hat{s}'}$ is the coset representative of the symmetric scalar manifold in
four-dimensions in the solvable parametrization, as a matrix in the adjoint representation of the solvable
group, $V_{4\,s}{}^{\hat{s}'}$ is the vielbein of the same manifold and the hat denotes rigid indices. \par The
conserved quantities are then obtained as the flux of the currents across the 2-sphere at infinity, according to
eq. (\ref{QA}):
\begin{align}
m&=\frac{1}{4\pi}\,\int_{S_2} {}^{*_3}
J_0\,\,;\,\,\,\ell=-\frac{1}{4\pi}\,\int_{S_2} {}^{*_3}
J_\bullet\,\,;\,\,\,
\Gamma^M=\frac{\sqrt{2}}{4\pi}\,\mathbb{C}^{MN}\int_{S_2} {}^{*_3}
J_N\,,\nonumber\\\Sigma_s &=\frac{1}{4\pi}\,\int_{S_2} {}^{*_3}
J_s\,.
\end{align}
The other conserved quantity characterizing the axisymmetric solution is the angular momentum $M_\varphi$
along the rotation axis $Z$. The expression of  the angular momentum in terms of a conserved current can be
found in standard textbooks (see for instance \cite{Wald} and \cite{Straumann}). Here we would like to give an
expression of it in terms of quantities which are intrinsic to the $D=3$ effective action: the Killing vector
field $\psi=\partial_\varphi$ and $J_\bullet$. To this end we start from the representation of $M_\varphi$ as the integral over
the sphere at infinity $S_2^\infty$ of a suitable 2-form, as given in \cite{Wald}:
\begin{align}
M_\varphi&=\frac{1}{16 \pi}\int_{S_2^\infty} J^{(2)}\,\,\,;\,\,\,\,
J^{(2)}\equiv\sqrt{g}\,\epsilon_{\mu\nu\rho\sigma}\,\nabla^\rho
\psi^\sigma\,dx^\mu\wedge dx^\nu\,.
\end{align}
The above integral can also be written in the form:
\begin{align}
M_\varphi&=\frac{1}{8\pi}\int_{S_2^\infty}
\sqrt{g}\,g^{\mu\,[t}\,\Gamma_{\mu \varphi}^{r]}\,d\theta d\varphi
=\frac{1}{8\pi}\int_{S_2^\infty}
\sqrt{g}\,g^{\mu\,[t}\,g^{r]\nu}\partial_{[\mu}
g_{\nu]\varphi}\,d\theta d\varphi= \,,\nonumber\\
&=\frac{1}{8\pi}\int_{S_2^\infty}
\sqrt{g_{(3)}}\,\left[\frac{1}{2}\,g_{(3)}^{rr}g_{(3)}^{\varphi\varphi}\left(\partial_r\omega_\varphi
g^{(3)}_{\varphi\varphi}-\omega_\varphi\, \partial_r
g^{(3)}_{\varphi\varphi}+e^{4\mathcal{\mathcal{U}}}\,\omega_\varphi^2\partial_r\omega_\varphi+\right.\right.\nonumber\\
&\left.\left.+4\omega_\varphi\,
 g^{(3)}_{\varphi\varphi}\,\partial_r\mathcal{\mathcal{U}}\right)\right]\,d\theta d\varphi\,.\label{integralS}
\end{align}
Using the asymptotic behavior of the metric for axisymmetric
solutions \cite{Straumann}:
\begin{align}
\omega_\varphi & =\frac{2 M_\varphi}{r}\sin^2(\theta)+
O\left(\frac{1}{r^2}\right)\,\,;\,\,\,
g^{(3)}_{rr}=1+O\left(\frac{1}{r^2}\right)\,\,;\,\,\,
g^{(3)}_{\theta\theta}=r^2\,\left(1+O\left(\frac{1}{r}\right)\right)\,,\nonumber\\
&
g^{(3)}_{\varphi\varphi}=r^2\,\sin^2(\theta)\,\left(1+O\left(\frac{1}{r}\right)\right)\,\,;
\,\,\,e^{2\mathcal{U}}=1-\frac{2m}{r}+O\left(\frac{1}{r^2}\right)\,,
\end{align}
we see that only the first two terms in the integral (\ref{integralS}) survive the asymptotic limit and yield
contributions which are both proportional to $M_\varphi$, the second term contributing twice the first
to the asymptotic limit. The first contribution in particular can be expressed in terms of $\psi,\,J_\bullet$,
so that we can write:
\begin{align}
M_\varphi&=-\frac{3}{8\pi}\,\int_{S_2^\infty}\psi_{[i}\,J_{\bullet
j]}\,dx^i\wedge
dx^j=-\frac{3}{4\pi}\,\int_{S_2^\infty}\psi_{[\theta}\,J_{\bullet
\varphi]}\,d\theta\,d\varphi=\nonumber\\
&=\frac{3}{8\pi}\,\int_{S_2^\infty}\psi_{\varphi}\,J_{\bullet
\theta}\,d\theta\,d\varphi\,,\label{mphi}
\end{align}
where $\psi_\varphi=g^{(3)}_{\varphi\varphi}$.
\subsubsection{$G_{(3)}$-invariant characterization of the angular momentum}
Let us define a new constant $\mathfrak{g}$-matrix as follows:
\begin{align}
Q_\psi &=-\frac{3}{8\pi}\,\int_{S_2^\infty}\psi_{[i}\,J_{
j]}\,dx^i\wedge dx^j=\frac{3}{8\pi}\,\int_{S_2^\infty}\psi_{\varphi}\,J_{
\theta}\,d\theta\,d\varphi \in \mathfrak{g}\,.
\end{align}
In the asymptotic limit $r\rightarrow \infty$ the components of $J_m$ have the following behavior:
\begin{equation}
J_r=\frac{Q}{r^2}+O\left(\frac{1}{r^3}\right)\,\,;\,\,\,J_\theta=\frac{Q_\psi}{r^2}\,\sin\theta+O\left(\frac{1}{r^3}\right)\,.
\end{equation}
%\paragraph{Global symmetry.}
According to the general formula (\ref{mphi}), the angular momentum can be written as:
\begin{equation}
M_\varphi=k \, {\rm Tr}(T_\bullet^\dagger\,Q_\psi)\,.
\end{equation}
 As pointed out earlier, $G_{(3)}$ is the global symmetry group of the three-dimensional effective theory.
  As an isometry group, its elements have a non-linear action on the coordinates:
\begin{equation}
g\in G_{(3)}\,:\,\,z^a\,\,\longrightarrow\,\,\,\, z_g^a=z_g^a(z)\,,
\end{equation}
where $z_g^a(z)$ are non-linear functions of the $z^a$, depending on the parameters of the transformation $g$.
The same transformation, being a global symmetry, maps a solution $\xi^a(x)$ into an other one of the same theory
$\xi_g^a(x)$. The asymptotic limit $r\rightarrow \infty$, for the scalar fields, defines  a single point $\xi_0=(\xi_0^a)$ on the scalar manifold:
\begin{equation}
\lim_{r\rightarrow \infty} \xi^a(x)=\xi_0^a\,.
\end{equation}
Since the action of $G_{(3)}$ on the scalar manifold is transitive, we can always map the point at infinity to
the origin $O(\xi_0^a\equiv 0)$. Once we fix $\xi_0=O$, we can only act on the solutions  by means of the stability group $H^*$ of the origin.

From the definition (\ref{cm}) we deduce the transformation property of the matrix $\mathcal{M}(z)$ under an isometry $g$:
\begin{equation}
\mathcal{M}(z)\,\,\longrightarrow\,\,\,\,\mathcal{M}(z_g)=g\,\mathcal{M}(z)\,g^\dagger\,,\label{cmt}
\end{equation}
where, with an abuse of notation, we have used the same symbol $g$ to denote the matrix form of $g$ in the representation
of $\mathcal{M}$. The $\mathfrak{g}$-valued current $J_m=J_m(\xi(x))$ therefore transforms under an isometry $g$ by conjugation:
\begin{equation}
J_m(\xi)\,\,\longrightarrow\,\,\,\,J_m(\xi_g)=(g^\dagger)^{-1}\,J_m(\xi)\,g^\dagger\,,
\end{equation}
and so do the $\mathfrak{g}$-valued constant matrices $Q$ and $Q_\psi$:
\begin{align}
Q(\xi)&\,\,\longrightarrow\,\,\,\,Q(\xi_g)=(g^\dagger)^{-1}\,Q(\xi)\,g^\dagger\,\,\,;\,\,\,\,\,
Q_\psi(\xi)\,\,\longrightarrow\,\,\,\,Q_\psi(\xi_g)=(g^\dagger)^{-1}\,Q_\psi(\xi)\,g^\dagger\,.
\end{align}
Generic axisymmetric stationary solutions are distinguished from the static ones by the following $G_{(3)}$-invariant property:
\begin{equation}
\mbox{axisymmetric solutions}\,\,\,\,\,\Rightarrow\,\,\,\,Q_\psi\neq 0.
\end{equation}
 In particular for solutions in the same $G_{(3)}$-orbit as the KN-Taub-NUT one, ${\rm Tr}(Q_\psi^2)\neq 0$.
 In the universal model originating from Einstein-Maxwell supergravity in four dimensions, see Sect. \ref{universal}, $G_{(3)}={\rm SU}(1,2)$, and  we can evaluate on the  KN-Taub-NUT solutions $Q$ and $Q_\psi$ explicitly. Using the covariant expression for the matrix $\mathcal{M}$ in terms of $U,V,W$, given in Appendix \ref{appe} and
eq.s (\ref{uvw}) introduced in Section \ref{universal} we find:
\begin{align}
Q &=\left(
\begin{array}{lll}
 0 & 0 & (m-i\,\ell) \\
 0 & 0 & - \,\frac{q+i p}{\sqrt{2}} \\
  (m+i\,\ell) & \frac{q-i p}{\sqrt{2}} & 0
\end{array}
\right)\,,\nonumber\\Q_\psi &=\alpha \,\left(\begin{matrix} 0 & 0 & (\ell+ i\,m) \cr
 0 & 0 & -i \,(q+i p)/\sqrt{2} \cr
 (\ell- i\,m) & -i \,(q-i p)/\sqrt{2}& 0\end{matrix}\right)\,.\label{QQpsi}
\end{align}
Then:
\begin{equation}
{\rm Tr}(Q^2)=\frac{2}{k}\,(m^2+\ell^2-\frac{p^2+q^2}{2})\,\,,\,\,\,{\rm Tr}(Q_\psi^2)=
\frac{2 \alpha^2}{k}\,(m^2+\ell^2-\frac{p^2+q^2}{2})\,,\label{eqq1}
\end{equation}
where $\alpha\equiv M_\varphi/m$ and $k=1$ in the  fundamental representation of ${\rm SU}(1,2)$,  so that
\begin{equation}
\left(\frac{M_\varphi}{m}\right)^2=\alpha^2=\frac{{\rm Tr}(Q_\psi^2)}{{\rm Tr}(Q^2)}\,.\label{eqq2}
\end{equation}
We wish to stress here that the  above formula, although derived in the universal model, holds in all supergravity theories admitting the KN-Taub-NUT solution.
This is a $G_{(3)}$-invariant characterization of the angular momentum, which holds for all solutions in the
same $G_{(3)}$-orbit as the KN-Taub-NUT one. Using this result, we can  write the extremality parameter in
a $G_{(3)}$-invariant fashion:
\begin{equation}
c^2= m^2+\ell^2-\frac{p^2+q^2}{2}-\alpha^2=\frac k2 {\rm Tr}(Q^2)-\frac{{\rm Tr}(Q_\psi^2)}{{\rm Tr}(Q^2)}\,,\label{eqq3'}
\end{equation}
so that the extremality condition becomes:
\begin{equation}
c^2=0\,\,\Leftrightarrow \,\,\,\,{\rm Tr}(Q^2)=\frac{2}{k}\,\frac{{\rm Tr}(Q_\psi^2)}{{\rm Tr}(Q^2)}\,,\label{eqq3}
\end{equation}
from which it is apparent that, as opposed to the static case, extremality does not imply  nilpotency of $Q$, as noted in \cite{Fre:2012im}. Eq. (\ref{eqq3}) provides a $G_{(3)}$-invariant characterization of extremality.
There is a class of extremal rotating solutions for which both sides of this equation vanish separately. These are the ``ergo-free'' (under-rotating) solutions constructed in \cite{Rasheed:1995zv,Larsen:1999pp,Astefanesei:2006dd} and further generalized in \cite{Bena:2009ev} within cubic supergravity models. Below we shall comment on some general $G_{(3)}$-invariant properties  of these solutions in terms of the matrices $Q$ and $Q_\psi$.
%From it we may hah
%derive the \emph{constants of motion} characterizing the solution at
%radial infinity:
%\begin{equation}
%Q_A\propto k\,{\rm Tr}\left(T_A^\dagger \, Q\right)\,,
%\end{equation}
%which consist in the ADM mass (for $T_A=T_0$), the NUT charge (for $T_A=T_\bullet$), the $D=4$ scalar charges
%(for $T_A=T_r$) and the electric-magnetic charges  (for $T_A=T_M$).
\subsubsection{Under-rotating solutions.} In \cite{Rasheed:1995zv,Larsen:1999pp,Astefanesei:2006dd} under-rotating solutions were constructed within the Kaluza-Klein theory originating from pure gravity in $D=5$, as a limit of a dilatonic rotating black hole. In order to perform a similar limit in the context of supergravity, we need to consider a model which is larger than the universal one, but which contains it as a consistent truncation. The simplest choice is the $\mathcal{N}=2$ $t^3$-model in four dimensions, which consists of supergravity coupled to one vector multiplet, whose complex scalar field $t$ parametrizes a special K\"ahler manifold with prepotential $\mathcal{F}(t)=t^3$.
Upon time-like reduction to $D=3$ we end up with an Euclidean sigma-model with target space ${\rm G}_{2(2)}/[{\rm SL}(2)\times {\rm SL}(2)]$ and global symmetry group $G_{(3)}={\rm G}_{2(2)}$. Extremal solutions to this model were studied in \cite{Gaiotto:2007ag,Kim:2010bf,Fre:2012im}.\par
 We shall not enter into the mathematical details of model but limit ourselves to illustrate the procedure for generating an extremal under-rotating solution from a non-extremal rotating one. The scalar fields originating from the $D=4$ vector fields are four $(\mathcal{Z}^M)=(\mathcal{Z}^0,\,\mathcal{Z}^1,\,\mathcal{Z}_0,\,\mathcal{Z}_1)$, parametrizing the solvable generators $(T_M)=(T_0,\,T_1,\,T^0,\,T^1)$. Adopting a suitable representation of ${\rm G}_{2(2)}$ for the generators (for example the fundamental real ${\bf 7}$ representation), we can consider two commuting generators
 of Harrison transformations:
 \begin{equation}
 K_0\equiv \frac{1}{2}\,(T_0+T_0^\dagger)\,\,;\,\,\,\, K_1\equiv \frac{1}{2}\,(T^1+T^{1\,\dagger})\,,
 \end{equation}
 and ``boost'' the Kerr solution with parameters $m,\,\alpha$ using the Harrison transformation:
  \begin{equation}
 \mathcal{O}\equiv e^{\log(\beta_1 m)\,K_0+\log(\beta_2 m)\,K_1}\,,
 \end{equation}
 The resulting solution is a non-extremal axion-dilaton rotating black hole with ADM-mass, electric-magnetic and scalar charges and angular momentum depending on the Kerr parameters $m,\,\alpha$ and encoded in the $\mathfrak{g}_{2(2)}$-valued matrices:
  \begin{equation}
  Q=\mathcal{O}^{-1}\,Q^{(K)}\,\mathcal{O}\,\,;\,\,\,Q_\psi=\mathcal{O}^{-1}\,Q_\psi^{(K)}\,\mathcal{O}\,,
  \end{equation}
  $Q^{(K)}$ and $Q_\psi^{(K)}$ being the matrices corresponding to the original Kerr solution.
  We shall give the complete solution elsewhere, focussing here only on the characteristic quantities at radial infinity. Redefining $\alpha=\Omega \,m=M_\varphi/m$, these quantities read:
 \begin{align}
 M_{ADM}&=\frac{1}{8} \left( m^2(\beta_1 +3 \beta_2 )+\frac{1}{\beta_1 }+\frac{3}{\beta_2 }\right)\,\,;\,\,\,p^1=\sqrt{3}\frac{m^2 \beta_2 ^2-1}{2 \sqrt{2} \beta_2 }\,\,;\,\,\,q_0=-\frac{M^2 \beta_1 ^2-1}{2 \sqrt{2} \beta_1 }\,,\nonumber\\
 \Sigma&=i\,\frac{\sqrt{3} \left(-m^2 \beta_2  \beta_1 ^2+m^2 \beta_2 ^2 \beta_1 +\beta_1 -\beta_2 \right)}{8 \beta_1  \beta_2 }\,\,;\,\,\,\,M_\varphi =\frac{\left(\beta_1  \beta_2 ^3 m^4+3 \beta_2  (\beta_1 +\beta_2 ) m^2+1\right) \Omega }{8 \sqrt{\beta_1 } \beta_2 ^{3/2}}\,,
 \end{align}
 while $p^0=q_1=\ell=0$.
 Taking the  the $m\rightarrow 0$ limit while keeping $\beta_1,\,\beta_2$ and $\Omega$ fixed, the above quantities remain finite:
  \begin{align}
 M_{ADM}&=\frac{1}{8} \left( \frac{1}{\beta_1 }+\frac{3}{\beta_2 }\right)\,\,;\,\,\,p^1=-\frac{\sqrt{3}}{2 \sqrt{2} \beta_2 }\,\,;\,\,\,q_0=\frac{1}{2 \sqrt{2} \beta_1 }\,\,;\,\,\,
 \Sigma=i\,\frac{\sqrt{3} \left(\beta_1 -\beta_2 \right)}{8 \beta_1  \beta_2 }\,\,;\,\,\,\,M_\varphi =\frac{\Omega }{8 \sqrt{\beta_1 } \beta_2 ^{3/2}}\,.
 \end{align}
 Inspection of the full solution shows that, as $m\rightarrow 0$, the ergo-sphere disappears and  the three dimensional spatial part of the metric becomes conformally flat.\par
This limit corresponds to taking a  singular Harrison transformation $\mathcal{O}$ ($\log(\beta_1\,m),\,\log(\beta_2\,m)\rightarrow -\infty$) and at the same time a singular limit of the Kerr parameters ($m,\,\alpha\rightarrow 0$). As a result the matrices $Q,\,Q_\psi$ remain finite but become \emph{nilpotent}. In particular $Q$ is a step-3 nilpotent matrix while $Q_\psi$ is step 2. The fact that $Q_\psi$ has a \emph{lower} degree of nilpotency than $Q$ is consistent with the fact that:
  \begin{equation}
  \lim_{m\rightarrow 0}{\rm Tr}(Q^2)=0\,\,;\,\,\,\, \lim_{m\rightarrow 0}\frac{{\rm Tr}(Q_\psi^2)}{{\rm Tr}(Q^2)}=0\,,
  \end{equation}
  and the extremality condition  (\ref{eqq3}) is satisfied. This is consistent with the classification of extremal solutions of \cite{Bossard:2012ge,Fre:2012im} in terms of suitable nilpotent subalgebras $\mathfrak{N}$ of $\mathfrak{g}$. In this case the matrices $Q$ and $Q_\psi$ would correspond to characteristic generators of $\mathfrak{N}$.

\subsection{A duality invariant expression for the DWHJ vector $S_m$}
Let us now apply the construction of section \ref{hj} to our specific  effective Lagrangian \eq{geolag}. The
direct application of eq. \eq{kda} to our specific geodesic model is possible but lacks the property of being
manifestly  invariant under the isometry group $G_{(3)}$. However, the use of the $G_{(3)}$-valued
matrix $\cM$ introduced in \eq{cm} makes it possible to write an alternative expression for $S^m$ which does
exhibit manifest duality invariance (provided we transform both the off-shell fields $z^a$ and their on-shell
expression on a given background $\xi^a(x)$). The expression is the following:
\begin{equation}
S^m=-\frac k{4 }\hat\rho\,\sqrt{h}\, {\rm Tr}\left[\mathcal{M}^{-1}(z)\partial^m \cM(\xi)  \right]
+
\delta^m_r \int^{r} {dr}^\prime \cL(\xi(x'),\partial_m\xi,x')\,.  \label{kdt}
\end{equation}
Indeed, from \eq{kdt} we find:
\begin{eqnarray}
\frac{\partial S^m}{\partial z^a} &=&\frac k{4 }\hat\rho\,\sqrt{h}\, {\rm Tr}\left[\mathcal{M}^{-1}(z)\frac{\partial\cM}
{\partial z^a}\mathcal{M}^{-1}(z) \partial^m \cM(\xi)  \right]\,,
\end{eqnarray}
so that, for a weakly embedded solution $z=\xi$, we reproduce the on-shell expression of the conjugate momentum
\eq{pigeo}. Correspondingly we also find, using the field equations:
\begin{equation}
\partial_m S^m|_{z=\xi} = \left(\cL  -\frac k{4 }\hat\rho\,\sqrt{h}\,
{\rm Tr}\left[\mathcal{M}^{-1}(z)\partial_m\cM (\xi)\mathcal{M}^{-1}(\xi) \partial^m \cM(\xi)
  \right]\right)_{z=\xi} = -\cH|_{z=\xi}\,.
\end{equation}
One may ask what the relation between the solution \eq{kdt} and the general relation \eq{kda} is. The answer can
be found by realizing that a Taylor-expansion of $S^m$ given in \eq{kdt} in powers of $z-\xi$, taking into
account \eq{cosetr3} and \eq{cm}, exactly reproduces \eq{kda}. It is important to stress that $S_m$,
as defined above, is $G_{(3)}$-invariant provided we simultaneously transform $z^a$ and $\xi^a(x)$
in its expression, as it follows from the transformation property (\ref{cmt}) of the matrix $\mathcal{M}$:
\begin{equation}
g\in G_{(3)}\,:\,\,\,S_m(z,\xi)\,\,\longrightarrow\,\,\,\,S_m(z_g,\xi_g)=S_m(z,\xi)\,,\label{Smt}
\end{equation}
An important property of the DWHJ construction is that one can compute the conserved currents of the theory by
varying $S^m$ with respect to the parameters which it depends on \cite{Kastrup:1982qq}. In particular, we can reproduce the conserved
N\"other currents $\hat \rho J_m$ of \eq{curr} by performing an infinitesimal isometry transformation on $S^m$,
at fixed background $\xi^a(x)$, and then by varying $S^m$ with the corresponding symmetry parameters. If we set:
\begin{equation}
g= \textbf{1} + \epsilon^\alpha \textbf{T}_\alpha
\end{equation}
the isometry transformed matrix is
\begin{equation}
\cM(z_g) = g\cdot \cM(z) \cdot g^\dagger\simeq \textbf{1} + \epsilon^\alpha\left( \textbf{T}_\alpha\cdot \cM +
\cM \cdot \textbf{T}_\alpha^\dagger\right)\,.
\end{equation}
On the $g$-transformed $S^m$ we get:
\begin{eqnarray}
\left.\frac{\partial S^m(z_g)}{\partial\epsilon^\alpha}\right\vert_{z=\xi} &=&-\frac k{4 } \hat \rho\,\sqrt{h}\,
\left[\left(\cM^{-1}(z) \partial^m \cM(\xi)\right)_i{}^j (T_\alpha)_j{}^i+\right.\nonumber\\
&&\left.+\left(\cM^{-1} (z)\partial^m \cM(\xi)\right)^j{}_i (T_\alpha)^i{}_j\right]
= -2 \,\hat \rho \,\sqrt{h}\,\, {\rm Tr}[T_\alpha^\dagger \cdot J^m]\,.\nonumber\\&&
\end{eqnarray}

%%%%%%%%%%%%%%%%%%%%%%%%%%%%%%%%%%%%%%%%%%%%%%%%%%%%%%%%%%%%%%%%%
%%%%%%%%%%%%%%%%%%%%%%%%%%%%%%%%%%%%%%%%%%%%%%%%%%%%%%%%%%%%%%%%%
\section{Application to Einstein--Maxwell axisymmetric solutions}\label{universal}

In the absence of four dimensional scalar fields ($\partial_i \phi = 0$, $\cM_{(4)}\to -\II$), the geodesic part
of the Lagrangian (\ref{geodaction}) reduces to
\begin{eqnarray} \label{3dnosc}
\frac{1}{\sqrt{g_{(3)}}}\,\mathcal{L}_{(3)}
&=&
\partial_i \cU \partial^i \cU+ \tfrac{1}{2}\e^{-2\,\cU}\,\partial_i{{\bf
Z}}^T\,\partial^i{{\bf Z}}
+ \tfrac{1}{4}\e^{-4\,\cU}\,(\partial_i{a}+{\bf Z}^T\mathbb{C}\partial_i{{\bf
Z}})(\partial^i{a}+{\bf Z}^T\mathbb{C}\partial^i{{\bf
Z}})\nn\\&=& \frac 12 G_{ab}(z)\partial_i z^a\,\partial^i z^b \,.
\end{eqnarray}
where $G_{ab}(z)$ is now the metric of the manifold:
\begin{equation}
\frac{{\rm SU}(1,2)}{{\rm U}(1)\times {\rm SU}(1,1)}\,,\label{pseudo}
\end{equation}
which is a pseudo-K\"ahler manifold, that is a non compact version of the K\"ahler manifold $CP(2)$.

 As it is well
known in General Relativity, a very simple and useful way to describe such theory is the use of the so-called
Ernst potentials $\cE$, $\Psi$ \cite{Chandrasekhar:1985kt,Stephani:2003tm} defined as:
\begin{equation}
\mathcal{E}=e^{2\cU}-|\Psi|^2+i\,a\,\,\,;\,\,\,\Psi=\frac1{\sqrt 2}(\mathcal{Z}^0+i\,\mathcal{Z}_0)\,,
\end{equation}
In terms of the Ernst potentials the metric (\ref{geo3}) reads:
\begin{equation}
dS^2_{(3)}=\frac{e^{-4\cU}}{2}\,|d\mathcal{E}+2\,\bar{\Psi}d\Psi|^2-2\,e^{-2\cU}\,|d\Psi|^2\,.
\end{equation}
The group $\rm{SU(1,2)}$ acts non-linearly on the potentials $\cE, \Psi$. However, one can introduce homogeneous
complex coordinate fields $(W,V,U)$ transforming in the ${\mathbf{3}}$ of $\rm{SU(1,2)}$, in terms of which the
Ernst potentials can be written as follows:
\begin{equation}
\cE = \frac{U -W }{U +W }\,;\quad \Psi =\frac{V }{U +W }
\end{equation}
Going to inhomogeneous variables  $u=U /W ,\,v=V/W$, they take the form
\begin{equation}
\mathcal{E}=\frac{u-1}{u+1}\,\,\,;\,\,\,\,\Psi=\frac{v}{u+1}\,.
\end{equation}
The scalar manifold $\frac{{\rm SU}(1,2)}{{\rm U}(1)\times {\rm SU}(1,1)}$ can then be described in terms of the
complex fields $z^a=(u,v)$ (where $a=1,2$).

We notice that the manifold (\ref{pseudo}) is a non-compact version of the minimal model $\frac{{\rm
SU}(1,2)}{{\rm U}(1)\times {\rm SU}(2)}$, which describes a particular case of a symmetric space of $N=2$
special geometry in four dimensional supergravity. Accordingly, we can say that the variables $(u,v)$
 are "special coordinates" in terms of which the upper components of the corresponding holomorphic symplectic
section $(X^\Lambda , F_\Lambda)$ read:
\begin{equation}
X^\Lambda=\left(\begin{matrix}W \cr V\cr U\end{matrix}\right)=W\,\left(\begin{matrix}1 \cr v\cr u\end{matrix}\right)\,,
\end{equation}
while the lower components $F_\Lambda$ are given in terms of the holomorphic homogeneous degree two prepotential
$F(X^\Lambda$), as $F_\Lambda = \frac {\partial F}{\partial X^\Lambda}$. The holomorphic prepotential in terms
of the inhomogeneous coordinates reads:
\begin{equation}
\mathcal{F}=\frac 1{W^2} F(X^\Lambda)=\frac{i}{4}\,(1-u^2-v^2)\,,
\end{equation}
and the K\"ahler potential $\mathcal{K}$ has the following form:
\begin{equation}
\mathcal{K}=
-\log\left[i\,\left(2\,(\mathcal{F}-\bar{\mathcal{F}})-(z^a-\bar{z}^a)(\mathcal{F}_a+
\bar{\mathcal{F}}_a)\right)\right]=-\log\left[|u|^2+|v|^2-1\right]\,.\label{kal}
\end{equation}

The coordinate patch $u,v$  is defined by the condition:
\begin{equation}
|u|^2+|v|^2>1.
\label{horcond}
\end{equation}
whose physical meaning will be given in the next subsection.\par
 The $\sigma$-model metric in the special coordinates has the form:
\begin{align}
dS_{(3)}^2&=2\,G_{a\bar{b}}\,dz^a\,d\bar{z}^b\,\,\,;\\
G_{a\bar{b}}&=\partial_a\partial_{\bar{b}}\mathcal{K}=e^{2\mathcal{K}}\,\left(\begin{matrix}(1-|v|^2) & \bar{u}\,v \cr \bar{v}\,u & (1-|u|^2)\end{matrix}\right)=e^{2\mathcal{K}}\,(\delta_{a\bar{b}}-z_a \bar{z}_{\bar{b}})\,,\label{kahlermet}\\
G^{\bar{a}b}&=-e^{-\mathcal{K}}\,(\delta^{\bar{a}b}-\bar{z}^{\bar{a}} z^b)\,.\nonumber
\end{align}
where $z_a\,\equiv \epsilon_{ab}\,z^b$
The eigenvalues of $g_{a\bar{b}}$ are: $-1/(|u|^2+|v|^2-1)\,,1/(|u|^2+|v|^2-1)^2$ and, if $|u|^2+|v|^2>1$, $g_{a\bar{b}}$
has the correct signature $(-,-,+,+)$.
\subsection{Relation to known black-hole solutions}\label{chan}
For stationary, axisymmetric, asymptotically flat solutions admitting the two
Killing vectors $\partial_t$ and $\partial_\varphi$, the most general case of complex
scalar fields $u,v$ corresponds to a Kerr--Newman solution with NUT-charge, whose metric reads
\cite{Stephani:2003tm}:
\begin{equation}
ds^2= \frac{\tilde\Delta}{|\rho|^2}(dt+\omega)^2-\frac{|\rho|^2}{\tilde\Delta}\left(\frac{\tilde\Delta}{\Delta} dr^2 + \tilde\Delta d\theta^2 +\Delta\sin^2\theta d\varphi^2\right)
\end{equation}
where
\begin{eqnarray}
\Delta &=& (r-m)^2 - c^2\,,\label{delta}\\
\tilde\Delta&=& \Delta -\alpha^2 \sin^2\theta\,,\label{deltatil}\\
\rho &=& r+ \ii\left(\alpha\cos\theta +\ell\right)\,,\label{rho}\\
\omega &=& \left(\alpha\sin^2\theta \frac{|\rho|^2-\tilde\Delta}{\tilde\Delta}+2\ell\cos(\theta)\right)d\varphi\,,\label{B}
\end{eqnarray}
where $c^2 = m^2+\ell^2 -\frac 12 (q^2+p^2) -\alpha^2$ as given in \eq{eqq3'}, in terms of  the Boyer--Lindquist coordinates $(r,\theta)$,
 of the electric and magnetic charges $(q,p)$ and of
the ADM-mass and NUT charge $(m,\ell)$. The parameter $\alpha$, as before, is related to the angular momentum $M_\varphi$ of
the solution by $\alpha= M_\varphi/m$. Here the metric field $\cU(r,\theta)$ is given by $e^{2\cU}=
\frac{\tilde\Delta}{|\rho|^2}$. For this solution the fields $\lambda,\,\hat{\rho}$ and the flat 2D metric $h_{mn}$ read:
\begin{align}
\lambda^2&=\tilde{\Delta}\,\,;\,\,\,\,\hat{\rho}=\sqrt{\Delta}\,\sin\theta\,\,;\,\,\,h_{mn}  \left(
      \begin{array}{cc}
        1/\Delta & 0 \\
        0 & 1 \\
      \end{array}
    \right)\,,
\end{align}
so that $\sqrt{h}\,\hat{\rho}=\sin(\theta)$. The latter expression holds, in suitable coordinates, for all axisymmetric solutions.
The Ernst potentials are then:
\begin{eqnarray}
\cE&=& \frac{r-2m +\ii (\alpha\cos\theta  - \ell)}{r +\ii (\alpha\cos\theta  + \ell)}\label{kntn1}\\
\Psi&=&\frac{-q +\ii p  }{\sqrt 2[r +\ii (\alpha\cos\theta  + \ell)]}\,.\label{kntn2}
\end{eqnarray}
 and the corresponding homogeneous coordinates can be chosen as:
\begin{eqnarray}
U&=&r-m +\ii \alpha\cos\theta\nn\\
V&=&\frac 1{\sqrt 2}(-q +\ii p)\nn\\
W&=&m+\ii \ell\label{uvw}
\end{eqnarray}
Let us observe that only an $SU(1,1)$ subset of the $SU(1,2)$ invariance is realized on the four dimensional
fields, under which the  "charges" $(W,V)$ form a doublet while $U$ is a singlet. The KN solution is retrieved
by setting $\ell=0$ in eq.s \eq{kntn1}, \eq{kntn2}, the RN electric-magnetic solution by further setting $\alpha
=0$ and finally the Schwarzschild solution is obtained from RN when $q=p=0$.

Let us relate the explicit expressions for the Ernst potentials here with the $\sigma$-model description given
above. The metric function $\tilde\Delta$ in \eq{deltatil} appears to be related to the $SU(1,2)$-invariant
K\"ahler potential $\cK$ in \eq{kal}:
\begin{equation}
\tilde\Delta = |U|^2+|V|^2-|W|^2= |W|^2 e^{-\cK}
\end{equation}
According to the identification \eq{uvw} the condition \eq{horcond} acquires a precise physical meaning.
 In the static solutions ($\alpha=0$) condition \eq{horcond} is guaranteed as long as $r>r_+$, $r_+$ being the outer horizon
\begin{equation}
r_+=m+\sqrt{m^2+ \ell^2-\frac{p^2+q^2}{2}}.
\end{equation}
On the other hand, in the KN case ($\ell=0$) it gives
\begin{equation}
r> m+\sqrt{m^2 -\frac{q^2+p^2}2-\alpha^2\cos^2\theta}\equiv r_e
\end{equation}
where $r_e>r_+ $ defines the external boundary of the ergosphere, where the component $g_{00}$ of the metric
vanishes, while $r_+ =m+\sqrt{m^2 -\frac{q^2+p^2}2-\alpha^2}$ is the radius of the outer event horizon. Then we
see that the special-coordinate patch described by $u,v$ is only valid outside the ergosphere.

If we cross the ergosphere surface $\tilde \Delta=0$ we are bound to change the coordinate patch. The new patch
can be described by the CP(2) riemannian space $SU(1,2)/U(2)$, with Kaehler potential $\cK =-\log (1-|u|^2
-|v|^2)$.\par The universal model considered here, and the KN-Taub-NUT solution thereof, can be embedded in more
general supergravity models (for instance in all $\mathcal{N}=2$ symmetric supergravity models, dimensionally reduced to
$D=3$) and thus it is interesting to consider  the $G_{(3)}$-invariant properties of this solution. In light of
the discussion at the end of  Sect \ref{axi},  the description of such properties should take into account, aside from  the
N\"other charge matrix $Q$, also the constant matrix $Q_\psi$.

\subsection{The DWHJ principal 1-form for the KN solution} Let us explicitly compute here the
DWHJ principal functions $ S^r,  S^\theta$ for the KN solution.

We have\footnote{We recall, from section \ref{chan}, that the two-dimensional metric is
\begin{equation}\label{d2metric}
  h_{mn} = \left(
      \begin{array}{cc}
        1/\Delta & 0 \\
        0 & 1 \\
      \end{array}
    \right)
\end{equation}}:
\begin{eqnarray}
\partial_{a} S^{m} &=&  \pi^{m}_{a}= \sin\theta\,  G_{a {\bar b}}(z ) h^{mn}\, \partial_n\bar{ z} ^{ {\bar b} }\label{expl}
\end{eqnarray}
that is:
 \begin{eqnarray}
\pi^{  r}_{a }&=& \sin\theta\,  G _{a  {\bar b} }(z ) \Delta\partial_r \bar{ z} ^{ {\bar b} }
\label{explr}\\
\pi^{  \theta}_{a }&=& \sin\theta\,  G _{a  {\bar b} }(z ) \bar\partial_\theta \bar{ z} ^{ {\bar b} }\,.
\label{expltheta}\end{eqnarray}
 Eq. \eq{expl}, recalling \eq{kda}, admits the (weakly embedded) solution:
\begin{equation}
S^{m}= 2 \Re \left[ \left(z^{ {a }}-\xi^{ {a }}(x)\right)\pi^{m}_{a }(x)\right] + \delta^m_r \int^r d\hat r\cL(\xi , \partial \xi , \hat x)
\end{equation}
 Using (\ref{kahlermet}), if we denote by $\xi^u, \xi^v$ the on-shell values of the fields $u ,v $:
 \begin{eqnarray}
 \xi^u &=& \frac{r-m+\ii\alpha\cos\theta}{m+\ii \ell}\nn\\
  \xi^v &=& \frac{-q+\ii p}{\sqrt 2(m+\ii \ell)}
  \end{eqnarray}
 we find
\begin{eqnarray}
S^{  r}(z ,x)&=&2\sin\theta\,(m^2+\ell^2)^2 \,\frac{\Delta (x)}{\tilde{\Delta}^2(x)}
\Re\left[(u-\xi^u) (1-|\xi^v)|^2)+(v-\xi^v) \xi^u\bar\xi^v\right]+\nn\\
&&+\int^r d\hat r\cL(\xi , \partial \xi , \hat x)\\
S^{  \theta}(z ,x)&=&-2\,\alpha\,\sin^2\theta\, \frac{(m^2+\ell^2)^2}{\tilde{\Delta}^2(x)}
\Im\left[(u-\xi^u) (1-|\xi^v)|^2)+(v-\xi^v) \xi^u\bar\xi^v\right]
\end{eqnarray}

%
%\begin{eqnarray}
%\label{esser1}
% \partial_{u }S _r & = & \gamma^{-1} \partial_{u}S_r\\
% \label{esser2} \partial_{v}S _r & = &  \partial_{v}S_r \\
%  \label{uequ} \partial_{a }S _\theta & = & \hat G_{a  \overline u }(\xi ) \bar \partial_\theta\bar u (u,\theta)\,.
%\end{eqnarray}

\section{Conclusions}\label{concl}
In this paper we have addressed the issue of the first order description of generic (not necessarily extremal) axisymmetric solutions. This was done by working out the general form of the principal functions $S_m$ associated with the corresponding effective 2D sigma-model in the DWHJ setting. We have also  given a characterization of the  general  properties of such solutions with respect to the global symmetry group of the effective 2D sigma-model which describes them. This was done by introducing, aside from the N\"other charge matrix, a further characteristic constant  matrix $Q_\psi$, in the Lie algebra of $G_{(3)}$, associated with the rotational motion of the black hole.\par
As a  direction for further investigation  it would be interesting to generalize this analysis to more general stationary solutions, including (non necessarily extremal) multicenter black holes. In this respect, as emphasized earlier, there is virtually no conceptual obstruction in generalizing the DWHJ construction and the general formula for $S_m$, which we have mainly used here within a 2D effective sigma-model, to the full 3D effective description of stationary solutions.  It would moreover be interesting to analyze the axisymmetric solutions to symmetric supergravities from the point of view of the \emph{integrability} of the corresponding effective 2D sigma-model, which we have not exploited here. This latter property being related to the presence in a gravity/supergravity theory, once dimensionally reduced to $D=2$, of an infinite dimensional global symmetry group, generalizing the Geroch group of pure Einstein gravity (see for instance \cite{Breitenlohner:1986um,Julia:1981wc}).

\section*{Acknowledgements}
We wish to thank P. Fr\`{e} for interesting discussions. This work was partially supported
by the Italian MIUR-PRIN contract 2009KHZKRX-007
''Symmetries of the Universe and of the Fundamental Interactions''.

\appendix
\section{The $\mathfrak{su}(1,2)$-Algebra} \label{appe}
 Let us choose the ${\rm SU}(1,2)$-invariant  and the $H^*={\rm U}(1,1)$-invariant metrics $\eta$
and $\bar{\eta}$, respectively, to be:
\begin{equation}
\eta={\rm diag}(-1,1,1)\,\,;\,\,\,\bar{\eta}={\rm diag}(-1,1,-1)\,,
\end{equation}
where the latter defines the coset generators. The solvable Lie algebra $Solv$ defining the Iwasawa
decomposition of $\mathfrak{su}(1,2)$ with respect to $\mathfrak{u}(2)$
 is generated by:
 \begin{align}
 Solv&={\rm span}(H_0,T_1,T_2,G)\,,\nonumber\\
 H_0&=\left(
\begin{array}{lll}
 0 & 0 & \frac{1}{2} \\
 0 & 0 & 0 \\
 \frac{1}{2} & 0 & 0
\end{array}
\right)\,\,;\,\,\,T_1=\left(
\begin{array}{lll}
 0 & -\frac{1}{2} & 0 \\
 -\frac{1}{2} & 0 & \frac{1}{2} \\
 0 & -\frac{1}{2} & 0
\end{array}
\right)\,\,;\,\,\,T_2=\left(
\begin{array}{lll}
 0 & -\frac{i}{2} & 0 \\
 \frac{i}{2} & 0 & -\frac{i}{2} \\
 0 & -\frac{i}{2} & 0
\end{array}
\right)\,,\nonumber\\
G&=\left(
\begin{array}{lll}
 -\frac{i}{2} & 0 & \frac{i}{2} \\
 0 & 0 & 0 \\
 -\frac{i}{2} & 0 & \frac{i}{2}
\end{array}
\right)\,.
 \end{align}
 The $H^*$ algebra $\mathfrak{u}(1,1)$ is generated by the compact component $K_\bullet$ of $G$, the non-compact components
 $K_1,\,K_2$ of $T_1,\,T_2$, respectively, and the compact $D=4$ duality generator $K$:
 \begin{align}
 \mathfrak{u}(1,1)&={\rm span}(K_1,K_2,K_\bullet,{K})\,,\nonumber\\
 K_\bullet&=G-G^\dagger= \left(
\begin{array}{lll}
 -i & 0 & 0 \\
 0 & 0 & 0 \\
 0 & 0 & i
\end{array}
\right)\,\,;\,\,\,K_1=T_1+T_1^\dagger=\left(
\begin{array}{lll}
 0 & -1 & 0 \\
 -1 & 0 & 0 \\
 0 & 0 & 0
\end{array}
\right)\,,\nonumber\\K_2&=T_2+T_2^\dagger=\left(
\begin{array}{lll}
 0 & -i & 0 \\
 i & 0 & 0 \\
 0 & 0 & 0
\end{array}
\right)\,\,;\,\,\,K=\left(
\begin{array}{lll}
 -i  & 0 & 0 \\
 0 & 2 i   & 0 \\
 0 & 0 & -i
\end{array}
\right)\,.
 \end{align}
 The ${\rm SU}(1,2)/{\rm U}(1,1)$-coset representative describing the physical patch of the manifold is:
 \begin{equation}
 \mathbb{L}=e^{-a G}\,e^{\sqrt{2}(\mathcal{Z}^0\,T_1+\mathcal{Z}_0\,T_2)}\,e^{2UH_0}\,.
 \end{equation}
 The matrix $\mathcal{M}=\mathbb{L}\bar{\eta}\mathbb{L}^\dagger$ has the following simple form:
 \begin{align}
 \mathcal{M}&=\mathbb{L}\bar{\eta}\mathbb{L}^\dagger=\eta-\frac{2}{I_2}\,\eta \overline{\mathbb{U}}\,\mathbb{U}^T\eta\,,
 \end{align}
 where
 \begin{equation}
 \mathbb{U}\equiv\left(\begin{matrix}W\cr V\cr U\end{matrix}\right)\,\,,\,\,\,I_2\equiv \mathbb{U}^T \eta \overline{\mathbb{U}}=|U|^2+|V|^2-|W|^2\,.
 \end{equation}

\section{KN Solution from Schwarzschild}\label{symcle}

In this appendix we give an alternative way  to generate the Hamilton principal 1-form $S^{(1)}$ corresponding
to the KN solution. It makes use of duality symmetry and general coordinate transformations starting from the
Schwarzschild solution.

We will proceed in two steps. We first need an explicitly $SU(1,2)$-duality invariant expression for the
$\cW_{3}$ of the RN solution in 3D. This can be achieved by using the generating technique of $SU(1,2)$ to
generate solutions in 3D. In particular, starting from Schwarzschild field variables
\begin{eqnarray}
U&=&r-m \nn\\
V&=&0\nn\\
W&=&m,
\end{eqnarray}
the action of the $SU(1,2)$  Harrison and Ehlers transformations generate electric, magnetic and in general also
a NUT charge, thus leading to a RN-NUT solution. Next,as a second step we use a procedure first introduced by
Cl\ ement \cite{Clement:1997tx} allowing the generation of a KN solution from RN by an appropriate sequence of
$SU(1,2)$ and coordinate transformations.

\subsection{$\mathcal{W}_3$ for the RN-NUT Solution}\label{symcle1}

Let us recall that in the static case  the prepotential $\mathcal{W}_3$ provides a first order description of
$D=3$ static solutions \cite{Andrianopoli:2009je}:
\begin{equation}
\frac{d\bar{ z}^{\bar{a}}}{d\tau}=g^{\bar{a}b}\,\partial_b\mathcal{W}_3\,\label{1ord}
\end{equation} satisfying the HJ equation

\begin{equation}
\partial_{\bar{a}}\mathcal{W}_3\,g^{\bar{a}b}\,\partial_b\mathcal{W}_3=c^2\,\label{HJeq}
\end{equation}
$c$ being the extremality parameter.\\ Quite generally a static solution is completely defined by a point $P$ of
the scalar manifold representing the values of the scalars at radial infinity $\tau=0$, and the  tangent vector
to the geodesic, which is an object transforming under $H^*$. Here $H^*$ is the isotropy group of the coset
$G/H^*$, $G$ being the 3D isometry group. Since the action of $G/H^*$ on $P$ is transitive over the scalar
manifold, we can always fix $P$ to be the origin $O$ at which all fields vanish, and study the geodesic
solutions corresponding to various choices of the velocity vector at infinity. In this way we break $G$ to the
little group $H^*$ of the origin and we expect the $\mathcal{W}_3$ describing the family of solutions with $P=O$
to be an $H^*$-invariant function.
\\In our case we have $G/H^* =\frac{{\rm SU}(1,2)}{{\rm U}(1)\times {\rm SU}(1,1)}$ and we shall prove that the
RN-NUT solutions are described by a solution to the HJ equation of the form:
\begin{equation}
\mathcal{W}_3=-c\,\log\left(\frac{|U|+\sqrt{|W|^2-|V|^2}}{|U|-\sqrt{|W|^2-|V|^2}}\right)=
-c\,\log\left(\frac{|u|+\sqrt{1-|v|^2}}{|u|-\sqrt{1-|v|^2}}\right)\,.\label{W3RNN}
\end{equation}
The above function is clearly $H^*={\rm U}(1,1)$-invariant since both $|U|$ and $|W|^2-|V|^2$ are.

Let us recover the expression \eq{W3RNN} for the $\mathcal{W}_3$ describing the most general static
(non-extremal) black hole in our model, from the one-parameter $\mathcal{W}^{(S)}_3$ of the Schwarzschild
solution by a \emph{duality (isometric) continuation} of it on the whole $\sigma$-model. By duality continuation
we mean \emph{defining} the value of $\mathcal{W}_3$ out of the one-dimensional submanifold on which
$\mathcal{W}^{(S)}_3$ is defined by means of an isometry transformation on the $\sigma$-model. Of course here we
are restricting to $H^*$ transformations only and the resulting prepotential will be, by construction,
$H^*$-invariant and still a solution to (\ref{HJeq}) being the latter duality invariant.
\par The geodesic corresponding to the Schwarzschild
black hole is defined by the following prepotential:
\begin{equation}
\mathcal{W}^{(S)}_3(s)=-c\log\left(\frac{s+1}{s-1}\right)\,,
\end{equation}
defined on the submanifold:
\begin{equation}
u=\bar{u}=s\,\,;\,\,\,\,v=0\,.
\end{equation}
It is straightforward to check that $\mathcal{W}^{(S)}_3(s)$ satisfies the HJ equation:
\begin{equation}
\partial_s \mathcal{W}^{(S)}_3 \frac{\partial s}{\partial \bar{z}^{\bar{a}}}\,g^{\bar{a} b}\,
\frac{\partial s}{\partial {z}^{{b}}}\partial_s \mathcal{W}^{(S)}_3=\frac{(s^2-1)^2}{4}\left(\partial_s \mathcal{W}^{(S)}_3\right)^2=c^2\,,
\end{equation}
where we have written $s=(u+\bar{u})/2$ and $z^a=(u,v)$. Next we apply to the Schwarzschild  fields a generic
$H^*$-transformation $h^*$. The latter can be written as the product of a Harrison transformation, a Ehlers
${\rm U}(1)_E$-transformation and a second ${\rm U}(1)$-transformation (which corresponds to the $D=4$ duality
group). Referring to the notations of Appendix \ref{appe} we have:
\begin{align}
h^*&=H_{arrison}\,\cdot h_E\,\cdot h\,\,\nonumber\\
H_{arrison}&=e^{a_1\,K_1+a_2\,K_2}=\left(\begin{matrix}\cosh({\bf a}) & -e^{i\sigma}\,\sinh({\bf a})& 0\cr
-e^{-i\sigma}\,\sinh({\bf a})& \cosh({\bf a}) & 0\cr 0&0&1\end{matrix}\right)\,,\nonumber\\
h_E&=e^{\alpha\,K_\bullet}={\rm diag}(e^{-i \alpha},1,e^{i \alpha})\,\,;\,\,\,h=e^{\beta\,K}={\rm diag}(e^{-i \beta},e^{2i \beta},e^{-i \beta})\,,
\end{align}
where we have written $a_1+i\,a_2={\bf a}\,e^{i\sigma}$. If we apply $h^*$ to the Schwarzschild  fields
described by $(W(s),V(s),U(s))=(1,0,s)$ we find:
\begin{equation}
\left(\begin{matrix}W\cr V\cr U\end{matrix}\right)=h^*\,\left(\begin{matrix}1\cr 0\cr s\end{matrix}\right)\,,
\end{equation}
that is:
\begin{equation}
u=\frac{U}{W}=e^{2i\alpha}\,\frac{s}{\cosh({\bf a})}\,\,;\,\,\,v=\frac{U}{W}=-e^{-i\sigma}\,\tanh({\bf a})\,.
\end{equation}
From the above relations we find $s$ in terms of the duality-transformed variables $u,v$:
\begin{equation}
s=\frac{|u|}{\sqrt{1-|v|^2}}\,.\label{s}
\end{equation}
Then we define $\mathcal{W}_3$ by duality continuation of $\mathcal{W}^{(S)}_3$:
\begin{equation}
\mathcal{W}^{(RN)}_3(u,v,\bar{u},\bar{v})=\mathcal{W}^{(S)}_3(s(u,v,\bar{u},\bar{v}))=-c\,\log\left(\frac{|u|+\sqrt{1-|v|^2}}{|u|-\sqrt{1-|v|^2}}\right)\,,
\label{SRN}\end{equation}
thus obtaining (\ref{W3RNN}).

We may explicitly check our result by solving the corresponding first order equations (\ref{1ord})
\begin{align}
\frac{d\bar{u}}{d\tau}&=c\bar{u}\,\frac{|u^2|-k^2}{|u|\,k}\,\,;\,\,\,k^2=1-|v|^2>0\,,\nonumber\\
\frac{dv}{d\tau}&=0\,.
\end{align}
From the first we derive:
\begin{equation}
\frac{d|u|}{d\tau}=c \,\frac{|u^2|-k^2}{k}\,\,\Rightarrow\,\,\,\,|u|=k\,\frac{A\,e^{2c\tau}+1}{1-A\,e^{2c\tau}}\,,
\end{equation}
where $A$ is an arbitrary constant that we take equal to 1. The second equation is telling us that $v$ also is
an arbitrary complex constant which we can set to:
\begin{equation}
v=-\frac{q-i p}{\sqrt{2} m}\,e^{i\alpha}\,\,\,\Rightarrow\,\,\,k=c/m\,.
\end{equation}
Being the phase of $u$ a constant, the general solution can be written as follows:
\begin{equation}
u=k\,\frac{e^{2c\tau}+1}{1-e^{2c\tau}}\,e^{2i\alpha}\,.
\end{equation}
Setting the arbitrary constant $A=0$ and using the relation between $\tau $ and $r$:
\begin{equation}\label{erretau}
\tau=\frac{1}{2c}\log\left(\frac{r-m-c}{r-m+c}\right)\,,
\end{equation}
we find:
\begin{equation}\label{RNvalues}
u=\frac{r-m}{m}\,e^{2i\alpha}\,\,\,;\,\,\,\,v=-\frac{q-i p}{\sqrt{2} m}\,e^{i\alpha}\,,
\end{equation}
which defines the RN-Taub-NUT solution where $m,p,q$ are the parameters of a RN solution and $\alpha$ is the effect
of a Ehlers ${\rm U}(1)$-transformation. The N\"other charge matrix reads:
\begin{equation}
Q=\frac 12\mathcal{M}^{-1}\frac{d}{d\tau}\mathcal{M}=\left(
\begin{array}{lll}
 0 & 0 &  e^{2i \alpha } m \\
 0 & 0 & -i e^{i \alpha } \frac{p-i q}{\sqrt{2}} \\
 e^{-2i \alpha }  m &  e^{-i \alpha }\frac{q-i p}{\sqrt{2}} & 0
\end{array}
\right)\,.\label{QRNN}
\end{equation}
The fields are obtained by the general formulas:
\begin{equation}
\cU=\frac{1}{2}\,\log\left(\frac{|u|^2+|v|^2-1}{|1+u|^2}\right)\,\,;\,\,\,
\Psi=\frac{v}{1+u}\,\,;\,\,\,a=-i\frac{u-\bar{u}}{|1+u|^2}\,.
\end{equation}
Using the generators of the solvable algebra of {$\frac{{\rm SU}(1,2)}{{\rm U}(1)\times {\rm SU}(1,1)}$ ( see
Appendix) we can compute the physical charges in terms of the parameters of the solution. The ADM mass $\hat{m}$ and NUT charge read:
\begin{equation}
\hat{m}={\rm Tr}(H_0^\dagger\,Q )=m\,\cos(2\alpha)\,\,;\,\,\,\,\ell=-\,{\rm Tr}(G^\dagger\,Q )=-m\,\sin(2\alpha)\,.
\end{equation}
while the complex charge $\frac{\hat{q}+i\,\hat{p}}{\sqrt 2}$ is:
\begin{equation}
\frac{\hat{q}+i\,\hat{p}}{\sqrt 2}=- \,{\rm Tr}((T_1+i T_2)^\dagger\,Q )=\frac{{q}+i\,{p}}{\sqrt 2}e^{i\alpha}\,.
\end{equation}
Using the above identifications, the matrix $Q$ in (\ref{QRNN}) reduces to the N\"other charge matrix in the first of  eq.s (\ref{QQpsi}), identifying hatted with un-hatted quantities. This represents the fact that the N\"other charge matrix $Q$ is the same for the KN-Taub-NUT and the RN-Taub-NUT solutions. The difference resides in the matrix $Q_\psi$ which vanishes in the latter solution.\par
Since the Maxwell-Einstein theory is a consistent truncation of a generic $\mathcal{N}=2$ model, the above procedure for constructing a manifestly $H^*$-invariant $\mathcal{W}_3$ for the generic solution in the same $G_{(3)}$-orbit as the Schwarzschild one, from  a \emph{duality completion} of $\mathcal{W}^{(S)}_3$, applies to a generic $\mathcal{N}=2,\, D=4$ supergravity. In this case the N\"other charge $Q$ of a generic representative of the Schwarzschild  orbit, is a diagonalizable matrix in the space $\mathfrak{K}$, orthogonal complement of $\mathfrak{H}^*$ in $\mathfrak{g}$ (the point at infinity $\xi_0$ is always set to coincide with the origin $O$), and transforms under the adjoint action  of $H^*$ in a characteristic $H^*$-representation. In particular $Q$ can be diagonalized using an $H^*$-transformation. The modulus $s$ in $\mathcal{W}^{(S)}_3$ is a function of the eigenvalues of $Q$, and thus is an $H^*$-invariant function of the parameters $Q_A$ of $Q$: $s=f(Q_A)$. These parameters also provide a parametrization of the coset $G_{(3)}/H^*\equiv e^\mathfrak{K}$ and, in the physical patch ${\Scr U}$, can be expressed in terms of the scalar fields $z^a$, so that we can locally express $s$ as a $H^*$-invariant function of $z^a$: $s=f(Q_A(z^a))=s(z^a)$. A duality completion procedure, analogous to the one illustrated above, allows then to determine the following $H^*$-invariant expression for $\mathcal{W}_3$ for the Schwarzschild  orbit:
\begin{equation}
\mathcal{W}_3=-c\log\left(\frac{s(z^a)+1}{s(z^a)-1}\right)\,.
\end{equation}
In the case of the universal model $s(z^a)$ was given in eq. (\ref{s}).

\subsection{The Cl\'ement Generating Technique}

Having at our disposal a duality invariant $\mathcal{W}_3$ for the RN solution, we may now apply a procedure,
introduced in \cite{Clement:1997tx}, to relate static and rotating black-hole solutions. In this way we shall
arrive at the explicit expression of the $U,V,W$ variables \eq{uvw} of the KN(-NUT) solution. We shall apply to
the RN set of homogeneous variables associated to (\ref{RNvalues}), which for definiteness we choose to be
\begin{equation}
U= r-m\,,\quad V=-\frac 1{\sqrt 2} (q-\ii p)\,,\quad W=m+\ii \ell
\label{uvwRN}
\end{equation}
 the transformation $\Pi\cdot R \cdot \Pi$, where:
\begin{equation}
\Pi: \{ U\to V, V\to U, W\to -W\}\label{ernani}
\end{equation}
is a $SU(1,2)$ involution, and $R$ is the following 4D space-time coordinate transformation:
\begin{eqnarray}
R: \left\{ \begin{array}{ccc}
              d\varphi &=& d\varphi' +\gamma \Omega d t' \\
             dt&=&\gamma dt'
           \end{array}\right.\,,
\end{eqnarray}
 relating the original reference frame to one rotating with constant angular velocity
 $\Omega$. The constant time-rescaling factor $\gamma$ will be fixed in the following to have the standard
 expression for the Ernst potentials of the KN solution.

 The first involution $\Pi$ gives rise to the following
new potentials:
\begin{eqnarray}
\cE'& =&\frac{U'-W'}{U'+W'}= \frac{-\frac 1{\sqrt 2} (q-\ii p) +m-i\ell}{-\frac 1{\sqrt 2} (q-\ii p) -m+i\ell}\,,\nonumber\\
\Psi'&=&\frac{V'}{U'+W'}= \frac{r-m}{-\frac 1{\sqrt 2} (q-\ii p) -m+i\ell}
\end{eqnarray}
One can readily see that the new solution corresponds to a Bertotti-Robinson space-time, with radius
$R_{BR}\equiv |V-W|= \sqrt{(\frac q{\sqrt 2}+m)^2+(\frac p{\sqrt 2}+\ell)^2}$ \cite{Clement:1997tx}.

The coordinate transformation $R$ induces the following transformation of the 4D static metric and gauge fields:
\begin{eqnarray}
 R: \left\{ \begin{array}{lll}
              e^{2\tilde \cU'} &=& \gamma^2 \left(e^{2\cU'} -e^{-2\cU'}\hat\rho^2\Omega^2\right) \\
             \tilde\omega&=& \frac{\hat\rho^2 \Omega}{\gamma\left( e^{4\cU'}-\hat\rho^2\Omega^2\right)}\\
             \tilde {\hat\rho}&=& \gamma\hat\rho
           \end{array}\right.
\end{eqnarray}
where
\begin{eqnarray}
e^{2\cU'}&=&\frac{|U|^2+ |V|^2-|W|^2}{R_{BR}^2}\equiv \frac\Delta{R_{BR}^2}\\
\tilde a'=a'&=& \frac{(\bar V W- V \bar W)}{R_{BR}^2} = \frac {2(e\ell-gm)}{R_{BR}^2}
\end{eqnarray}
We have introduced here the $SU(1,2)$ invariant $\tilde \Delta$, which, in the coordinates \eq{uvwRN}, is:
\begin{equation}
\tilde\Delta= (r-m)^2 -c_{RT}^2
\end{equation}
where $c^2_{RT}\equiv |W|^2-|V|^2= m^2+\ell^2-\frac 12(q^2+p^2)$ is the extremality parameter of the dyonic
RN-NUT solution. Note that $c^2_{RT}= \frac k2 \,\mathrm{Tr}[Q^2]$ (see eq. \eq{eqq1}).

 The redefinition of the metric implies a transformation of the gauge field-strengths, that corresponds to the
following transformation on the gradient of the Ernst potential $\Psi$ (here $x^m=(r,\theta)$):
\begin{eqnarray}
\partial_m \tilde\Psi'= \gamma\left[\partial_m \Psi' -\hat\rho \Omega e^{-2\cU'}\,({}^{\star_{(2)}} \partial_m {\overline{\Psi'}})\right]\,.\label{dpsir}
\end{eqnarray}
The integration of eq. \eq{dpsir} is easily performed by observing that ${}^{\star_{(2)}} \partial_r
 {\overline{\Psi'}}=0$ since $\Psi'=\Psi'(r)$ is only function of the radial variable. Further observing that
$\partial_r\Psi'= -\frac\gamma{R^2_{BR}}\left[(e+m) +\ii (\ell+g)\right]$, the final result is
\begin{eqnarray}
\tilde\Psi'&=& \gamma\{\Psi'(r) +\ii(V-W)\Omega \cos\theta\}\nn\\
&=& \frac\gamma{R^2_{BR}} \left\{ (r-m)(\bar V-\bar W) + \ii \alpha\cos\theta\right\}
\end{eqnarray}
together with
\begin{eqnarray}
\tilde\cE'&=& e^{2\tilde \cU'} -|\tilde\Psi'|^2 +\ii \tilde a'
\nn\\
&=& -\frac{\gamma^2}{R^2_{BR}}\left(c^2_{RT}+\alpha^2\right)+\frac{\ii\,(\bar V W- V \bar W)}{R_{BR}^2}
\end{eqnarray}
where we have defined $\alpha\equiv (\Omega R^2_{BR})$.

We may give a simpler expression to the Ernst potentials by fixing the time rescaling $\gamma$ as
\begin{equation}
\gamma^2 =\frac{c^2_{RT}}{c^2_{RT} + \alpha^2}\,.
\end{equation}
With this redefinition we obtain
\begin{eqnarray}
\tilde\cE'&=& \frac{\tilde U'-\tilde W'}{\tilde U'+\tilde W'} =\frac{V+W}{V-W}\\
\tilde\Psi'&=&\frac{\tilde V'}{\tilde U'+\tilde W'} =\frac{\gamma(U+\ii \alpha\cos\theta)}{V-W}\,.
\end{eqnarray}
implying the following transformation on the homogeneous variables:
\begin{equation}
R\cdot\Pi:
\left\{ \begin{array}{rcl}
              \tilde U' &=& V \\
             \tilde V'&=& \gamma(U+\ii \alpha\cos\theta) \\
             \tilde W'&=& -W
           \end{array}\right.
\end{equation}
Performing again the transformation $\Pi$ as given in \eq{ernani}, we finally obtain the KN (TaubNUT) fields in
terms of the corresponding variables of the RN (TaubNUT) solution:
\begin{equation}
\Pi\cdot R\cdot\Pi:
\left\{ \begin{array}{rcl}
              \tilde U'' &=& \gamma(U+\ii \alpha\cos\theta) \\
             \tilde V''&=& V \\
             \tilde W''&=& W
           \end{array}\right. \label{prp}
\end{equation}
corresponding to the potentials
\begin{eqnarray}
\tilde\cE''&=&  =\frac{\gamma(U+\ii\alpha\cos\theta) -W}{\gamma(U+\ii\alpha\cos\theta)+W}\\
\tilde\Psi''&=& =\frac{V}{\gamma(U+\ii\alpha\cos\theta)+W}\,.
\end{eqnarray}
They coincide with the standard KN potentials (see, for example, \cite{Stephani:2003tm}, Chapter 21)
\begin{eqnarray}
\cE_{KN}&=&  1-\frac{ 2m  }{r+\ii\alpha\cos\theta}\\
\Psi_{KN}&=& \frac{-\frac 1{\sqrt 2} (q-\ii p)}{r+\ii\alpha\cos\theta}\,.
\end{eqnarray} if we set, besides $\ell=0$:
\begin{equation}
r\to \gamma(r-m)+m\,,\quad \alpha \to \gamma\alpha\,.
\end{equation}
For the KN solution, the field $a$ appearing in \eq{geodaction} is given by the imaginary part of $\cE$,
\begin{equation}
a=2 \frac{ m\alpha \cos\theta}{|\rho|^2}
\end{equation}

\end{document}